\documentclass[12pt]{article}
\usepackage{epsfig}
\usepackage{hyperref}
\usepackage{amssymb}
\begin{document}

\title
{                   
\begin{flushright}{\normalsize HU-EP-09/60 \\
                           LU-ITP 2009/002} 
\end  {flushright}\vspace{5mm}
Two-point functions of quenched lattice QCD
in Numerical Stochastic Perturbation Theory.
\\
(I) The ghost propagator in Landau gauge}

\author
{F.~Di~Renzo  \\
\small
{Dipartimento di Fisica, Universit\`a di Parma and INFN, I-43100 Parma, Italy}
\and E.-M.~Ilgenfritz  \\
\small
{Institut f\"ur Physik, Humboldt-Universit\"at zu Berlin, D-12489 Berlin, Germany}
\and
H.~Perlt, A.~Schiller  \\
\small
{Institut f\"ur Theoretische Physik, Universit\"at Leipzig, D-04009 Leipzig, Germany}
\and
C.~Torrero  \\
\small
{Institut f\"ur Theoretische Physik, Universit\"at Regensburg, D-93040 Regensburg, Germany}
}

\maketitle

\begin{abstract}
This is the first of a series of two papers on the perturbative computation 
of the ghost and gluon propagators in $SU(3)$ Lattice Gauge Theory. Our final aim is to 
eventually compare with results from lattice simulations in order to enlight the 
genuinely non-perturbative content of the latter. By means of Numerical Stochastic 
Perturbation Theory we compute the ghost propagator in Landau gauge up to three loops. 
We present results in the infinite volume and $a \to 0$ limits, based on a general 
strategy that we discuss in detail. 
\end{abstract}

\section{Introduction}
\label{sec:introduction}

The identification of confinement mechanisms is a challenge for Lattice 
QCD~\cite{Greensite:2003bk,Alkofer:2006fu,Greensite:2007zz}.
This requires to single out observables which both capture the relevant physical 
content and are in the end suitable for numerical simulations, so that a 
non-perturbative investigation becomes possible. 
The gluon and the ghost propagators with their momentum dependence are believed 
to encode specific features and mechanisms of a 
confining theory~\cite{Alkofer:2006fu}, which can be studied once a gauge is fixed. 
Apart from the non-unique asymptotic infrared behavior, which -- if
calculated in lattice simulations -- realizes the so-called decoupling
solution instead of the scaling solution 
(see Refs.~\cite{vonSmekal:2008ws,Fischer:2008yv,Bogolubsky:2009dc})
of the Schwinger-Dyson and Functional Renormalization 
Group~\cite{Fischer:2008uz} approaches~\footnote{The latter solution 
is preferred in view of the Gribov-Zwanziger~\cite{Gribov:1977wm,Zwanziger:1991gz} 
and Kugo-Ojima~\cite{Ojima:1978hy,Kugo:1979gm} confinement scenarios.
The conflict with lattice calculations has given rise to a controversy 
about the (in)dispensability of having a BRST-invariant lattice gauge-fixing
prescription. It is argued to be possible~\cite{vonSmekal:2008es} although 
still lacking. This interesting development is probably irrelevant for the 
aspects discussed below.}, in these functions the onset of non-perturbative 
effects is generally agreed to be associated with the following two aspects:
the coupling to local condensates~\cite{Boucaud:2008gn} and to localized vacuum excitations like
vortices~\cite{Langfeld:2001cz,Gattnar:2004bf}.
In order to effectively face the genuine non-perturbative effects, 
it is desirable to have for reference an understanding of the perturbative 
behavior of these gauge-dependent functions in 
higher than one-loop order.

Such a program cannot evade a well-known 
practical problem: Lattice Perturbation Theory (LPT) is much more involved compared to 
its continuum QCD counterpart. The complexity of the diagrammatic approaches
rapidly increases beyond the one-loop approximation~\cite{LPT}. 
By now, only a limited number of results up to two-loop accuracy has been obtained. 
As an effective alternative to diagrammatic LPT a numerical tool has been 
proposed by the Parma group in~\cite{Di Renzo:1994sy}. The method is a quite 
direct numerical implementation of Stochastic Perturbation Theory, which in 
turn comes as one of the two main applications of the ideas of Stochastic 
Quantization, the other being the use of Langevin dynamics as an alternative to
standard Monte Carlo techniques~\cite{Parisi:1980ys,Batrouni:1985jn}. 
This numerical tool for LPT goes under the name of Numerical Stochastic Perturbation Theory (NSPT) 
and has been reviewed in~\cite{Di Renzo:2004ge}. 

In a joined effort we started a new field of application for NSPT based on our
respective interests in LPT (Leipzig), NSPT (Parma) and Monte Carlo studies
of gluon and ghost propagators (Berlin).
First results have been presented 
at Lattice 2007~\cite{Ilgenfritz:2007qj,DiRenzo:2007qf}, Lattice 2008~\cite{DiRenzo:2008ir},
Confinement8~\cite{DiRenzo:2008nv} and at Lattice 2009~\cite{DiRenzo:2009ei}. 
The present paper is the first of a series of two. While this paper is concentrated 
on the ghost propagator, the forthcoming one will present the results for 
the gluon propagator. 

One of our main goals is to provide results for the comparison of 
higher-loop calculations of perturbative contributions with the results of
lattice Monte Carlo calculations of the main propagators in QCD on finite 
size lattices. The other goal consists in determining higher loop contributions 
to the logarithmically divergent perturbative ghost propagator in momentum space 
including non-logarithmic contributions to good precision. 
The present work also aims at a further check on the precision of
the method by comparison to known one-loop results from LPT.
In order to proceed to two- and higher-loop results in the infinite volume limit, 
peculiar emphasis will be put on the control of finite size effects. 
We want to stress that our results have been obtained from two completely 
independent sources: while strategies have been extensively 
discussed, the actual implementation of Parma and Leipzig NSPT codes are 
completely independent of each other, thus providing a stronger check on 
numerical results. 

The paper is organized as follows. 
In Sect. \ref{sec:Langevinequation} we present selected elements of NSPT which are 
relevant for the case at hand, in particular the issue of performing a NSPT computation in 
a fixed gauge, {\em i.e.} the minimal Landau gauge. 
Sect. \ref{sec:ghostpropagator} focuses on the observables we 
will be concerned with, while Sect. \ref{sec:standard-LPT} summarizes
known analytic results for the perturbative ghost propagator
up to three-loop order and their relation to the unknown
non-logarithmic coefficients we want to determine.
In the following section we present the technical details of 
practicing NSPT. Finally we collect our results in Sect.~\ref{sec:results} and 
briefly discuss a comparison to Monte Carlo data.

\section{Landau gauge fixing in NSPT}
\label{sec:Langevinequation}

The basis of Stochastic Quantization is the (stochastic) time evolution 
of fields driven by the Langevin equation. In the case of a $SU(N)$ lattice 
gauge theory the Langevin equation for the $SU(N)$ matrix $U_{x,\mu}$ located 
at the link between lattice sites $x$ and $x+\hat\mu$ reads
\begin{equation}
  \frac{\partial}{\partial t} U_{x,\mu}(t;\eta) = {\rm i}\; \bigl(
  \nabla_{x,\mu} S_G[U]- \eta_{x,\mu}(t) \bigr) \;
  U_{x,\mu}(t;\eta) \, ,
  \label{eq:Langevin_I}
\end{equation}
$t$ is the (stochastic) Langevin time. The so-called drift term is 
given by the equations of motion derived from the Euclidean gauge action 
$S_G[U]$: they are written in terms of the differential operator 
$\nabla_{x,\mu} = \sum_a~T^a\nabla_{x,\mu}^a$, 
which is a left Lie derivative. 
The $T^a$ ($a=1,\dots, N^2-1$) are the generators in the fundamental representation.
In our case we will have $N=3$ and the gauge action $S_G[U]$ is 
the Wilson plaquette action.

The key issue of Stochastic Quantization is that in the limit of large $t$ 
expectation values with respect to the noise converge to expectation values 
with respect to the QFT functional integral, {\em i.e.} 
the distribution of the gauge fields converges to the Gibbs measure
$P[U] \propto {\rm{exp}}(- S_G[U])$.
In any numerical approach (and this applies to our case), however, the Langevin equation 
is not solved for continuous time. One needs to discretize time as a sequence 
$t= n \epsilon$, with running step number $n$. Therefore, in order to extract 
correct physical information, one needs to perform the double 
extrapolation $n \to \infty$ and $\epsilon \to 0$. For the numerical solution 
of the Langevin equation we adhere to a peculiar version of the Euler scheme 
that guarantees all the link matrices $U_{x,\mu} \in SU(3)$ 
to stay in the group manifold:
\begin{eqnarray}
  U_{x,\mu}(n+1; \eta) & = & \exp \bigl( {\rm i}~F_{x,\mu}[U,\eta] \bigr) \; U_{x,\mu}(n; \eta)\ ,
  \label{eq:iteration}
  \\
  F_{x,\mu}[U, \eta] & = & \epsilon~\!\nabla_{x,\mu} S_G[U] + \sqrt{\epsilon}~\eta_{x,\mu} 
  \, .
  \label{eq:force}
\end{eqnarray} 
In matrix form and at finite $\epsilon$, the components of $\eta$ are subject to the mutual
correlations ($i,k,l,m=1,\dots,3$)
\begin{equation}
  \left\langle 
  \eta_{i,k}(z) \; \eta_{l,m}(w)
  \right\rangle_\eta
  =\left[
  \delta_{il} \delta_{km}- \frac{1}{3} \delta_{ik} \delta_{lm}
  \right]\delta_{zw}  
  \,.
  \label{eq:matrixform}
\end{equation}
$z$ and $w$ (integer) denote here Langevin time steps. The noise is white,
meaning that different time steps are uncorrelated.

The usual path from Stochastic Quantization to Stochastic Perturbation Theory 
would go through trading the differential form of Langevin equation for its 
integral equation counterpart in a Green function formalism, which is easy to 
solve in a perturbative scheme. NSPT remains instead in a differential formulation. 
For the case of Lattice Gauge Theories we rewrite each link matrix as an expansion 
in the bare coupling constant $g_0$. Since $\beta=2N/g_0^2$, the expansion reads
\begin{equation}
  U_{x,\mu}(t; \eta) \to 1 + \sum_{l \ge 1} \beta^{-l/2} U_{x,\mu}^{(l)}(t; \eta) 
  \,.
  \label{eq:expansion_of_U}
\end{equation}
Provided one rescales the time step to 
$\varepsilon = \beta \epsilon$, the expansion converts the Langevin 
equation (\ref{eq:iteration}) into a system of simultaneous 
updates in terms of the expansion coefficients (\ref{eq:expansion_of_U})  
of $U_{x,\mu}$ and of similar expansion coefficients for the force 
$F_{x,\mu}$ in Eq. (\ref{eq:force}). While the random noise $\eta$ enters 
only the lowest order equation, higher orders are stochastic by the noise 
propagating from lower order to higher order 
terms. The system can be truncated according to the maximal order of the gauge 
field one is interested in.

In a notation which is common in Lattice Perturbation Theory we can relate the 
(antihermitean) vector potential $A_{x+\hat{\mu}/2,\mu}$ to the link matrices $U_{x\mu}$ via
\begin{equation}
 A_{x+\hat{\mu}/2,\mu}  =  \log U_{x,\mu} \, .
\label{eq:log_mapping}
\end{equation}
An expansion similar to (\ref{eq:expansion_of_U}) exists for that potential 
taking values in the algebra $su(3)$,
\begin{equation}
  A_{x+\hat{\mu}/2,\mu}(t; \eta) \to \sum_{l \ge 1} \beta^{-l/2} 
  A_{x+\hat{\mu}/2,\mu}^{(l)}(t; \eta) \, .
  \label{eq:expansion_of_A}
\end{equation}
We observe that unitarity of $U_{x,\mu}$ has a counterpart in the antihermiticity and tracelessness 
of all orders of $A_{x+\hat{\mu}/2,\mu}$:
\begin{equation}
  A^{(l)\dagger}_{x+\hat{\mu}/2,\mu} = - A^{(l)}_{x+\hat{\mu}/2,\mu}\,, \quad 
  {\rm{Tr}}~A^{(l)}_{x+\hat{\mu}/2,\mu}=0 \, . 
\end{equation}
As we will see, the definition of the ghost propagator will in a sense be based on the 
expansion in the algebra. 

We finally remark that whenever we speak about contributions of some order to an observable this has
to be understood as an expansion
\begin{equation}
  \langle {\cal O} \rangle \to \sum_{l \ge 0} \beta^{-l/2} \langle {\cal O}^{(l)} \rangle \ , 
  \label{eq:expansion_of_O}
\end{equation}
and constructed by comparison to coefficients of $\beta^{-l/2}$. 

We now come to the main point of how to perform a gauge-fixed NSPT
computation, in the case at hand in Landau gauge. A key issue of the standard  
(in other words, non-perturbative) Langevin equation for LGT is the non-necessity of 
fixing a gauge. 
In short: the drift term in Langevin equation is equal to the lattice equations of motion 
(put equal zero there). As a consequence, longitudinal degrees of freedom undergo 
an unbounded diffusion, but the group manifold is compact
and hence no divergence shows up. Such a situation is no longer true in NSPT. This is quite 
obvious in the algebra: the diffusion of the longitudinal component of the $A_{\mu}$ fields 
is unbounded and hence their norms diverge in stochastic time.
We stress that gauge-invariant quantities are in principle not affected by these divergences, 
but this is no advantage: since norms are diverging one 
runs eventually into trouble by lack of accuracy. We also remark that this is true regardless of 
the choice of either expansion (\ref{eq:expansion_of_A}) or (\ref{eq:expansion_of_U}): 
perturbation theory implies decompactification in any case.

A solution comes along the lines suggested in~\cite{Zwa}: {\em fixing} (to some precision) 
{\em the gauge is going to fix the problem} and this goes under the name of Stochastic Gauge Fixing. 
We adhere here to the recipe of~\cite{Di Renzo:2004ge}. The Langevin evolution steps are 
interwoven with gauge transformations
\begin{equation}
  U_{x,\mu}^G = G(x) \, U_{x,\mu} \, G^\dagger(x + \hat \mu) \,.
  \label{eq:gtrafo}
\end{equation}
The main requirement for $G$ is quite obvious: gauge transformations should be expanded in 
the coupling as well and the leading order should not affect the background vacuum 
around which the solution to Langevin equation is expanded (the trivial vacuum, in the case at hand). 
A convenient solution comes with the choice
\begin{equation}
  G(x)= \exp \left\{ - \alpha \, \left( \sum_\mu \partial_\mu^L A_\mu \right)(x)\, \right\}  \,.
\label{eq:localstep}
\end{equation}

In general the action of the left/right lattice derivative $\partial^{L/R}_\mu$ 
on a function $f(x)$ defined on site $x$ is given by
\begin{equation}
  \partial_\mu^{L/R} f(x)= \pm \frac{1}{a} \left( f(x)-f(x \mp  \hat \mu) \right) \,.
  \label{eq:leftrightgradient}
\end{equation}
The action of the left lattice derivative on the vector potential 
defined on midpoints of the links is defined as
\begin{equation}
  \left( \sum_\mu \partial_\mu^L A^{(l)}_\mu \right)(x) = \frac{1}{a}
  \sum_\mu \left( A^{(l)}_{x+\hat\mu/2,\mu}-A^{(l)}_{x-\hat\mu /2,\mu} \right) 
  \,.
  \label{eq:leftgradientA}
\end{equation}
The resulting divergence is mapped to site $x$.
We will come back to this choice: the iteration of this gauge transformation would 
drive the field configuration to the Landau gauge. 

With respect to Stochastic Gauge Fixing, it suffices here to point out that one step of 
(\ref{eq:gtrafo}) using (\ref{eq:localstep}) interwoven with 
the Langevin step (\ref{eq:iteration}) is enough to keep fluctuations under control
provided that $\alpha$ is chosen of order $\alpha \sim \varepsilon$.
The rationale for this is the essence of Stochastic Gauge Fixing: any new drift term in 
the Langevin equation which amounts to a local gauge transformation does 
not affect the asymptotic distribution of gauge invariant quantities, 
while it can provide a restoring force in the 
non-gauge-invariant sector. The differential (finite difference, actually) 
operator present in (\ref{eq:localstep}) makes it clear that zero modes are completely 
unaffected up to this point. Their divergences are still present. 
We take the simplest prescription of subtracting zero modes at every order.

One can actually do more than this and perform a gauge-fixed NSPT computation 
in the case of covariant gauges. The interested reader is referred 
to~\cite{NSPT_FeynGauge}. The key 
issue is that one can rewrite the functional integral in the Faddeev-Popov formalism, which 
merely results in a different form of the gauge action (an effective action, one would say). 
Without entering into details, in this context it is worth making two points. First, there is 
no need for the introduction of ghost fields: one simply faces an extra contribution to the action 
in the form $S_{FP} = {\rm Tr} \,{\rm log} \, M$, $M$ being the Faddeev-Popov operator we will be 
concerned with in the following. Second, for the 
Landau gauge the procedure is of little help, since the 
other extra piece in the action is the gauge fixing action 
$S_{gf} = 1/(2\xi) \sum_x {\rm Tr} (\partial \cdot A)^2$, 
with $\xi$ being the covariant gauge parameter, which is zero in this case. 

The failure of the general covariant gauge NSPT procedure has anyway a counterpart: 
as a limit case $\xi \to 0$ of covariant gauges, the Landau gauge 
\begin{equation}
  \left( \sum_\mu \partial_\mu^L A^{(l)}_\mu \right)(x) = 0 \ ,
  \label{eq:landaucond}
\end{equation}
is related to an extremum of the norm of the field, and there is a well-known iterative procedure 
to reach such an extremum. The basic step is (\ref{eq:localstep}), {\em i.e.} the 
gauge transformation that we also use for Stochastic Gauge Fixing. The iteration stops when, 
for all orders $l$, the summed-up quadratic violations of 
(\ref{eq:landaucond}) vanish within double    
precision ($V$ is the lattice volume),
\begin{equation}
  \frac{1}{V} \sum_x {\rm Tr} \sum_{i=1}^{l-1}\left[
  \left(\sum_\mu \partial_\mu^L A^{(i)}_\mu \right)^\dagger (x)
  \left( \sum_\mu \partial_\mu^L A^{(l-i)}_\mu \right)(x) \right] = 0 \, . 
  \label{eq:condition}
\end{equation}
Iterating 
(\ref{eq:gtrafo}) with (\ref{eq:localstep})
is actually not very efficient, since it 
suffers from critical slowing down due to the low modes, which 
becomes more and more severe as the lattice size increases.

An effective solution to this problem is also 
well-known and goes through Fourier acceleration as introduced in $\!$~\cite{Davies:1987vs}. 
In NSPT one makes use of the expanded variant of the gauge transformation
\begin{equation}
  G(x) =\exp \left\{ \hat F^{-1} \left[ \alpha \,
  \frac{\hat p^2_{\rm{max}}}{\hat p^2} \,
  \hat F \left[ \left( \sum_\mu \partial_\mu^L A_\mu \right)(x) \right](p)\right] \right\} 
  \label{eq:fourieracc}
\end{equation}
in order to achieve the minimal Landau gauge.
$\hat{F}$ denotes the Fourier transformation from discrete $x$ to momentum
$p$ space and $\hat{F}^{-1}$ its inverse.
The gauge transformation (\ref{eq:fourieracc}) applied locally is actually 
not local in the sense of (\ref{eq:localstep}). 
The driving force, the divergence of the gauge potential,
is non-locally smeared with a low-pass filter that attenuates 
the higher Fourier components of $\partial_\mu^L A^{(l)}_\mu$ by a factor 
$\hat{p}^2_{\rm max}/\hat{p}^2$. Here $\hat{p}^2$ is a non-zero eigenvalue of 
the free field lattice Laplacian, and its maximal value on a periodic lattice 
in all directions is $a^2 \hat{p}^2_{\rm{max}}=16$. 
While we notice that the volume-independent 
optimal parameter $\alpha$ for the case of periodic boundary conditions in all 
directions is found for this perturbative Landau gauge fixing to be
$\alpha = 1/(a^2 \hat p^2_{\rm{max}})$, we also point out that Fourier acceleration also 
works if one takes the plain lattice momenta (which in the following we will reference 
as $p_\mu$) instead of $\hat{p}_\mu$. 

This is a good point to pin down some notations for lattice momenta.
In the following we will reserve the latin letter $k$ to denote $k_{\mu}$, 
the integer 4-tuples $(k_1,k_2,k_3,k_4)$, 
while plain lattice momenta will be denoted by $p_\mu(k_\mu)$
\begin{equation}
  p_{\mu}(k_\mu) = \frac{2 \pi k_{\mu}}{L_{\mu}} \,.
  \label{eq:plainMOM}
\end{equation}
Here $L_{\mu}$ is the lattice size in direction $\mu$ and the reference to lattice spacing
$a$ can be made explicit: $L_{\mu} = N_{\mu} a$.
In our case lattices will be symmetric, {\em i.e.} $L_{\mu}=L$ and $N_{\mu}=N$.
According to the two most popular conventions, 
$k_{\mu} \in \left(-N_{\mu}/2, N_{\mu}/2\right]$ or $k_{\mu} \in \left[0, N_{\mu}-1\right]$. 
The eigenvalues of the free Laplacian are 
\begin{equation}
  \hat{p}^2 = \sum_{\mu} [\hat{p}_\mu(k_\mu)]^2 \, ,
\end{equation}
where for each direction separately 
\begin{equation}
  \hat{p}_\mu(k_\mu) = \frac{2}{a} \sin\left(\frac{\pi k_{\mu}}{N_{\mu}}\right) = 
  \frac{2}{a} \sin\left(\frac{ a p_\mu}{2}\right) \, .
  \label{eq:p-definition}
\end{equation}

\section{The lattice ghost propagator}
\label{sec:ghostpropagator}

The ghost propagator $G$ is nothing but the inverse of the 
Faddeev-Popov operator $M$. 
We start our discussion in the continuum, where -- in the case 
of covariant gauges -- the Faddeev-Popov (FP)
operator reads
\begin{equation}
  M = - \nabla \cdot D(A) \, .
\end{equation}
$D(A)$ is the covariant derivative, emerging from the response of the gauge field 
$A_{\mu}$ to an infinitesimal gauge transformation. 
$G$ is diagonal in the (adjoint) color space ($a,b,\dots=1,\dots, 8$) 
\begin{equation}
  G^{ab}(p) = \delta^{ab} G(p^2) \, . 
\label{eq:diagonal}
\end{equation}       
As an average over field configurations ${\{A\}}$, it is translationally invariant
\begin{equation}
  G^{ab}(x-y)=G(x-y) \, \delta^{ab} = \left\langle {M^{-1}}^{ab}_{xy} 
  \right\rangle_{\{A\}} \, .
\end{equation}
In Landau gauge, because $\nabla \cdot A \equiv \sum_\mu \partial_\mu A_{\mu}(x)=0$, 
we can reverse the order of $\nabla$ and $D$, 
\begin{equation}   
  M= - \nabla \cdot D(A) = -D(A) \cdot \nabla \, .
\end{equation}

To obtain the form of the FP operator on the lattice one closely mimics the steps 
one goes through in the continuum. The lattice form of that operator is nevertheless 
much more involved, since the response of the Lie-algebra $A_{\mu}$ field 
to a gauge transformation does not have a closed form. 
In perturbation theory there is no logical obstruction, anyway: 
at any given order the lattice counterpart of the covariant derivative can be 
pinned down~\cite{Rothe,Torrero_thesis} and all in all the FP operator reads 
\begin{equation}
  M^{ab}_{xy}= \left[\partial_\mu^L {\cal D}_\mu(U)\right]_{ab} \delta_{x,y} \, .
\label{eq:FPop}
\end{equation}
Here we denote by ${\cal D}(U)$ the lattice counterpart of the covariant derivative, 
indicating a dependence on $U$, even if this dependence is through a perturbative expansion. 
This results in a dependence on the adjoint representation of the Lie-algebra field
\begin{equation}
  {\cal D}_\mu[\phi]= \left[1+ \frac{\rm i}{2} \Phi_\mu(x) - \frac{1}{12} \Phi_\mu^2(x) 
  - \frac{1}{720} \Phi_\mu^4(x) + \frac{1}{30240} \Phi_\mu^6(x) + \dots\right] \partial_\mu^R 
  + {\rm i}\,\Phi_\mu(x) \, .
\end{equation}
The $\Phi_\mu(x)$ is the adjoint representation of the lattice gauge field 
$\phi$ ($\phi \equiv {\rm -i} A$, $A$ antihermitean)
\begin{equation}
 \left[\Phi_\mu(x)\right]_{bc} = \left[t^a\right]_{bc} \phi_\mu^a\equiv -{\rm i} \, 
 f^{abc} \phi_\mu^a  \, ,
\end{equation}
the fundamental representation being given by ($i,j=1,2,3$):
\begin{equation}
  \left[\phi_\mu(x)\right]_{ij} = \left[T^a\right]_{ij} \phi_\mu^a\,, \quad
  \left[A_\mu(x)\right]_{ij} = \left[T^a\right]_{ij} A_\mu^a \, .
\end{equation}

The ghost propagator in momentum space can in principle be obtained 
by Fourier transformation in each lattice configuration ${\{U\}}$,
taking the translational invariance into account: 
\begin{equation}
  G^{ab}(p(k))= \frac{1}{V} \left\langle \; 
  \sum_{x,y} [M^{-1}]^{ab}_{xy} \; \exp\left({\rm i} \;p(k)\cdot(x-y)\right)
  \; \right\rangle_{\{U\}} = \left\langle [M^{-1}]^{ab}(p(k)) \right\rangle_{\{U\}}
  \label{eq:PFmom}\ ,
\end{equation}
with the scalar product $p(k) \cdot x = \sum_\mu 2 \pi k_\mu x_\mu /L_\mu$. 
Given the color-diagonal form, one is interested in the scalar function defined by 
the adjoint trace
\begin{equation}
  G( p(k))= \frac{1}{8} G^{aa}(p(k))=
    \frac{1}{8} \left\langle {\rm{Tr}_{adj}} M^{-1}(p(k))\right\rangle_{\{U\}} \, .
  \label{eq:trace}
\end{equation}
Due to its eight trivial zero eigenvalues (corresponding to global gauge
transformations) the inverse $M^{-1}$ needs to be defined with care.
Since one is interested in non-zero momenta $p(k)$, 
this means that one is automatically restricted
to a subspace orthogonal to the space spanned by the (space-time 
constant) zero modes of the Faddeev-Popov operator.

For an individual configuration, the inverse of the FP operator 
is not translationally invariant and hence (\ref{eq:PFmom}) is not so useful. 
The way out is quite standard and goes through the inversion of $M$ on a 
delta source in momentum space, {\it i.e.} a plane wave source 
$\psi(a,k)^b_x=\delta^{ab} \exp \left( {\rm i}\; p(k) \cdot x \right)$ 
in real space:
\begin{equation}
  [M^{-1}(k)]^{ab} =  \sum_{cd,xy} \psi(a,k)^{*,c}_x \, [M^{-1}]_{xy}^{cd} \, \psi(b,k)^d_y \,.
\end{equation}
As in standard non-perturbative applications, the issue is inverting the 
relevant operator on a source, but things are actually a lot easier in NSPT. 
The matrix $M$ can simply be considered as expanded using the expansion of the 
gauge fields $A^{(l)}$:
\begin{equation}
  M = M^{(0)} + \sum_{l>0} \beta^{-l/2} M^{(l)} \, .
\end{equation}
Then the inverse of $M$ will also obtained as a power series
\begin{equation}
  M^{-1} = [M^{-1}]^{(0)} + \sum_{l>0} \beta^{-l/2} [M^{-1}]^{(l)} \, . 
\end{equation}
One easily finds~\cite{Di Renzo:2004ge} a recursive way to obtain the higher orders of $M^{-1}$:
\begin{eqnarray}
  [M^{-1}]^{(0)}&=&[M^{(0)}]^{-1} \,,
  \\
  \nonumber 
  [M^{-1}]^{(l)}&=& - [M^{-1}]^{(0)}\;\sum_{j=0}^{l-1} M^{(l-j)}\;[M^{-1}]^{(j)} \, .
\end{eqnarray}
It is worth pointing out that the computation proceeds back and forth from 
position to momentum space, since at any step the operators at hand are diagonal 
(or almost diagonal) in one of the two spaces. 

This is the time to reintroduce perturbative (loop) indices: we denote the $n^{th}$ loop 
order of the ghost propagator as
\begin{equation}
  G^{(n)}(p(k)) = \left\langle[M^{-1}(p(k))]\right\rangle^{(l=2n)} \, .
\end{equation}
Thus a real loop order $n$ of the ghost propagator corresponds to 
contributions with even powers ${(\sqrt\beta)}^{-2l}$, 
whereas contributions with odd powers ${(\sqrt\beta)}^{-2l+1}$ would indicate 
a non-loop contribution which does not exist by construction in standard perturbation theory.
Contributions related to these non-integer $n$ have to vanish numerically in NSPT 
for sufficiently large statistics. 

Similar to the gluon propagator in the forthcoming paper, we present the various orders of
the ghost dressing function (or ``form factor'')
in two forms:
\begin{equation}
  J^{(n)}(p) = p^2 \; G^{(n)}(p(k)) \,, \quad
  \hat J^{(n)}(p) = \hat p^2 \; G^{(n)}(p(k)) \, .
\label{eq:dressingfunction}
\end{equation}
The dressing function has to be calculated for each 4-momentum
given  by four integers -- the momentum 4-tuple
$(k_1,k_2,k_3,k_4)$.

Two comments are in order for (\ref{eq:dressingfunction}). First of all,
the $J$'s are dimensionless quantities. Notice that in a numerical lattice
simulation one does not have access to $G(p(k))$, but to the dimensionless
quantity $a^{-2} G(p(k))$. Multiplying by $(ap)^2$ - or $a^2 \hat p^2$ - one obtains 
the corresponding (still dimensionless) $J(a,p)$ - or $\hat J(a,p)$ - we will be 
concerned with in the following.
Second, the second 
notation in (\ref{eq:dressingfunction}) is chosen to remind that here the quantities 
$\hat p_\mu$ are acting as the ``physical'' momenta.

\section{Relation to standard Lattice Perturbation Theory}
\label{sec:standard-LPT}

In the RI'-MOM scheme, the renormalized ghost dressing function $J^{\rm RI'}$ is 
defined as
\begin{equation}
  J^{\rm RI'}( p, \mu , \alpha_{\rm RI'}) = 
  \frac{J(a, p, \alpha_{\rm RI'})}{Z(a,\mu,\alpha_{\rm RI'})} \, , 
  \label{eq:renorm}
\end{equation}
with the condition
\begin{equation}
  J^{\rm RI'}( p, \mu , \alpha_{\rm RI'})|_{p^2=\mu^2} = 1 \, . 
  \label{eq:RIcondition}
\end{equation}
Therefore, the ghost dressing function $J(a,p,\alpha_{\rm RI'})$ is just the ghost 
wave function renormalization constant
$Z(a,\mu,\alpha_{\rm RI'})$ at $\mu^2=p^2$. 

The general form of an operator/function $O(a,\mu, \alpha)$ in terms of the 
renormalized coupling 
$\alpha(a\mu)= (g(a\mu))^2/(16\pi^2)$~\footnote{Note the additional factor $1/(4\pi)$ 
compared to the ``standard'' definition.} can be written as 
\begin{equation}
  O(a,\mu, \alpha)=1+\sum_{i>0} \alpha(a \mu)^i \,\sum_{j=0}^i\, O_{i,j}\,(\log(a\mu))^j \, .
  \label{eq:Opexpansion}
\end{equation}
Here $j$ gives the possible powers of the logarithms in loop order $i$.
                                                                                         
Using (\ref{eq:renorm}) and (\ref{eq:RIcondition}), we get the expansions
\begin{eqnarray}
  Z(a,\mu, \alpha_{\rm RI'})&=&1+\sum_{i>0} \alpha_{\rm RI'}^i 
  \,\sum_{j=0}^i\, z^{\rm RI'}_{i,j}\,\left(\log (a\mu)\right)^j \, ,
  \\
  J(a,p, \alpha_{\rm RI'})&=&1+\sum_{i>0} \alpha_{\rm RI'}^i
  \,\sum_{j=0}^i\, z^{\rm RI'}_{i,j}\,
  \left(\frac{1}{2}\log(pa)^2 \right)^j\,,
  \\
  J^{\rm RI'}(p,\mu, \alpha_{\rm RI'})&=&1+\sum_{i>0} \alpha_{\rm RI'}^i 
  \,\sum_{j=0}^i\, z^{\rm RI'}_{i,j}\,
  \left(\frac{1}{2}\log \frac{p^2}{\mu^2}\right)^j\,.
  \label{eq:expansion_of_J}
\end{eqnarray}
Since we measure the dressing function as an expansion in the lattice 
(inverse) coupling $\beta$, we  need the ghost wave function renormalization 
as an expansion in the lattice coupling
$ \alpha_0={N_c}/{(8 \pi^2 \beta)} $:
\begin{equation}
  Z(a,\mu,\alpha_0)=1+ \sum_{i>0} \alpha_0^i \, \sum_{j=0}^i\, 
  z_{i,j}\,\left(\log(a\mu)\right)^j\,.
  \label{Zcbare}
\end{equation}

{}From now we restrict ourselves to three-loop expressions in Landau gauge and in
the quenched approximation. The logarithmic corrections are partly known from the 
anomalous dimension of the ghost field and the beta function and given as follows 
(compare {\em e.g.} \cite{Gracey:2003yr})
\begin{eqnarray}
  &&
  z^{\rm RI'}_{1,1}= - \frac{3}{2} \, N_c \,,
  \\
  &&
  z^{\rm RI'}_{2,2}= - \frac{35}{8} \, N_c^2 \,, \quad
  z^{\rm RI'}_{2,1}= - \frac{271}{24}\, N_c^2 +
  \frac{35}{6} \, z^{\rm RI'}_{1,0} \,N_c  \,, 
  \\
  &&
  z^{\rm RI'}_{3,3}= - \frac{2765}{144} \, N_c^3 \,, \quad
  z^{\rm RI'}_{3,2}= - \frac{11933}{144} \, N_c^3
  + \frac{2765}{72} \, z^{\rm RI'}_{1,0}\, N_c^2
  \,, 
  \\
  &&
  z^{\rm RI'}_{3,1}=- \frac{157303}{864} \, N_c^3
  + \frac{211}{16} \, \zeta [3] \, N_c^3 
  + \frac{91}{8} \, z^{\rm RI'}_{1,0}\, N_c^2
  + \frac{79}{6} \, z^{\rm RI'}_{2,0}\, N_c\,. 
  \nonumber
\end{eqnarray}
The finite one-loop constant $z^{\rm RI'}_{1,0}$ has been calculated from the 
ghost self-energy in standard infinite volume LPT~\cite{Kawai:1980ja} with the value
\begin{equation}
  z^{\rm RI'}_{1,0}=13.8257 \, .
\end{equation}
To our knowledge, the constants $z^{\rm RI'}_{2,0}$ and $z^{\rm RI'}_{3,0}$ 
were not known so far.

Taking into account the relation between the lattice coupling and the renormalized 
coupling in two-loop accuracy~\cite{Hasenfratz:1980kn,Luscher:1995np,Christou:1998ws}
(the function $d_1$ is scheme independent)
\begin{equation}
  \alpha_{\rm RI'}= \alpha_{\rm {\overline{MS}}}+ O(\alpha_{\rm {\overline{MS}}}^5)=
  \alpha_0 + d_1 \, \alpha_0^2 + d_2 \, \alpha_0^3 + \dots \ ,
\end{equation}
with
\begin{eqnarray}
  &d_1(a,\mu) = - 2 \beta_0 \log (a\mu) +K_1 \,, &
  \\
  &d_2(a,\mu)=  \left(2 \beta_0 \log (a\mu)\right)^2 -2 (2 \beta_0 \, K_1 + 
  \beta_1) \log (a\mu) +K_2 \,, &
  \\
 & \beta_0 = \frac{11}{3} \, N_c   \,, \quad  
  \beta_1 = \frac{34}{3 }\, N_c^2 \,,  \quad
   K_1 =  - \frac{2 \pi^2}{N_c} + 26.8384 \, N_c \,,&
  \nonumber \\
  &K_2=K_1^2 + \frac{ 6 \pi^4}{N_c^2} -452.047 + 197.252 \, N_c^2  \,, &
\end{eqnarray}
we get the coefficients $z_{i,j}$ of $Z(a,\mu,\alpha_0)$ in (\ref{Zcbare}) up to three loops
\begin{eqnarray}
  &&
  z_{1,1}=z^{\rm RI'}_{1,1} \,, \quad z_{1,0}=z^{\rm RI'}_{1,0}\,, 
  \\
  &&  z_{2,2}=z^{\rm RI'}_{2,2}- 2 \beta_0 \,z^{\rm RI'}_{1,1} \,,\quad 
  z_{2,1}=z^{\rm RI'}_{2,1} + K_1 \, z^{\rm RI'}_{1,1}
                 - 2 \beta_0 \, z^{\rm RI'}_{1,0} \,,
  \\
  &&  z_{2,0}=z_{2,0}^{\rm RI'} + K_1 \, z^{\rm RI'}_{1,0} \,,
  \nonumber 
  \\
  &&  z_{3,3}=z^{\rm RI'}_{3,3}- 4 \beta_0 \,z^{\rm RI'}_{2,2}
  + 4 \beta_0^2 \,z^{\rm RI'}_{1,1} \,,
  \nonumber \\
  &&  z_{3,2}=z^{\rm RI'}_{3,2}+ 2 K_1 \,z^{\rm RI'}_{2,2} 
  - 4 \beta_0 \,z^{\rm RI'}_{2,1}- 4 \beta_0 \,K_1 \,z^{\rm RI'}_{1,1}
  + 4 \beta_0^2 \,z^{\rm RI'}_{1,0}- 2 \beta_1 \, z^{\rm RI'}_{1,1} \,,
  \nonumber  
  \\
  &&  z_{3,1}=z^{\rm RI'}_{3,1}+ 2 K_1 \,z^{\rm RI'}_{2,1} 
  +  K_2 \,z^{\rm RI'}_{1,1}
  - 4 \beta_0 \,z^{\rm RI'}_{2,0}- 4 \beta_0 \, K_1\,z^{\rm RI'}_{1,0}
  - 2 \beta_1 \, z^{\rm RI'}_{1,0} \,,
  \\
  &&  z_{3,0}=z^{\rm RI'}_{3,0}+  2 K_1 \,z^{\rm RI'}_{2,0}
  +  K_2 \,z^{\rm RI'}_{1,0}\,.
  \nonumber
  \label{d2}
\end{eqnarray}

Using these numbers and the inverse lattice coupling $\beta$ we obtain the 
expression for the ghost dressing function to that order
\begin{eqnarray}
  &&J^{\rm 3-loop}(a,p, \beta) =
  1+\sum_{i=1}^{3} \frac{1}{\beta^{i}} J^{(i)}(a,p) \,, \quad 
J^{(i)}(a,p)=\sum_{j=0}^i\, J_{i,j}\,
  \left(\log (pa)^2 \right)^j
  \label{Zghostbeta3loop}
  \,,
\end{eqnarray}
with
the one-loop coefficients
\begin{eqnarray}
  \label{J1loop}
  J_{1,1}&=&-0.0854897\,, 
  \nonumber
  \\ 
  J_{1,0}&=&0.0379954  \,z^{\rm RI'}_{1,0}=0.525314 \, ,
\end{eqnarray}
the two-loop coefficients
\begin{eqnarray}
  \label{J2loop}
  J_{2,2}&=&0.0215195 \,, 
  \nonumber \\
  J_{2,1}&=&-0.313514 - 0.00324822 \, z^{\rm RI'}_{1,0}=-0.358423\,,
  \\
  J_{2,0}&=&0.106737 \, z^{\rm RI'}_{1,0} + 0.00144365  \, z^{\rm RI'}_{2,0}
  =1.47572 + 0.00144365 \, z^{\rm RI'}_{2,0} \,, 
  \nonumber
\end{eqnarray}
and the three-loop coefficients
\begin{eqnarray}
  \label{J3loop}
  J_{3,3}&=&-0.00660927\,, 
  \nonumber \\
  J_{3,2}&=&0.164130 + 0.000817642\, z^{\rm RI'}_{1,0}=0.175434\,,
  \nonumber \\ 
  J_{3,1}&=&-1.38120 - 0.0210370\, z^{\rm RI'}_{1,0}- 0.000123418\, z^{\rm RI'}_{2,0}
  \nonumber \\
  &=&-1.67205 - 0.000123418\, z^{\rm RI'}_{2,0}
  =-1.54589 - 0.0854897 J_{2,0} \,,
  \\
  J_{3,0}&=&0.375990\, z^{\rm RI'}_{1,0}+0.00811105\, z^{\rm RI'}_{2,0}
  +0.0000548523\, z^{\rm RI'}_{3,0}
  \nonumber \\
  &=&5.19833 + 0.00811105\, z^{\rm RI'}_{2,0} +0.0000548523\, z^{\rm RI'}_{3,0}\,.
  \nonumber
\end{eqnarray}
Thus we conclude:
the coefficients of the leading logarithmic contributions for a given order can be exclusively
taken from continuum computations whereas non-leading log coefficients are already influenced
by finite lattice constants from corresponding lower loop orders.

In general, these logarithmic terms are the relatively
easy part of a diagrammatic perturbative computation, while getting the 
finite constants is the big effort. In NSPT it is just the other way around. 
As it will be clear in Section~\ref{subsec:fitting}, telling which is which in the 
measured ghost dressing function at each loop  
requires a fitting procedure: fitting a logarithmic term would actually 
require a terrific precision. 
As a consequence, we have just stated from the very beginning 
the (modest) attitude of not even trying to extract anomalous dimensions from 
our NSPT computations.
On the contrary, we have to assume that the coefficients in front of the logarithms
are known.
Assuming that a continuum calculation has been performed up to a given loop order
defines the maximum  accuracy and the maximal loop number
which we can reach in a lattice NSPT computation. 

We have to check that NSPT in the limits $\hat p a\to 0$ (or $ p a\to 0$) and $V\to\infty$ 
reproduces the constant $J_{1,0}$ from the one-loop result.
After that we may proceed to determine  from the two-loop measurements 
the unknown $J_{2,0}$ and, therefore, $z^{\rm RI'}_{2,0}$.
Knowing that number, we are in the position to fix the coefficients of all non-leading logarithmic 
contributions in the three-loop expression and to estimate
the constant $J_{3,0}$ or $z^{\rm RI'}_{3,0}$ from the three-loop measurements.

\section{The practice of NSPT simulations} 
\label{sec:practice}

\subsection{Statistics and extrapolation to zero Langevin step}
\label{subsec:statistics}

Choosing a maximal loop expansion order $n_{\rm {max}}= l_{\rm {max}}/2$ and 
solving the coupled system of equations for the gauge links $U^{(l)}_{x,\mu}$ and 
the gauge potentials $A^{(l)}_{x+\hat{\mu}/2,\mu}$ simultaneously, 
one generates a sequence of these fields for all perturbative orders 
$1, \dots,l _{\rm {max}}$. 
Such sequences --- representing Langevin time trajectories --- 
are generated for different finite Langevin time steps $\varepsilon$ and different
lattices sizes, {\em i.e.} different $N$.

{}From the perturbative fields at finite $\varepsilon$ we construct the observable of interest, 
in our case the ghost dressing function for each chosen set of lattice momenta and average
it over the used configurations within a Langevin time trajectory.
It is expected that the autocorrelation time  for the observable is extending over 
subsequent configurations and increases with decreasing $\varepsilon$.
As a reasonable compromise between computer time and autocorrelation we 
have measured the ghost propagator using a sequence of configurations separated 
by 50 Langevin steps.
Before measuring, Landau gauge has to be obtained on each configuration by
satisfying (\ref{eq:condition}) to high precision.
To further increase the statistics, we have generated several different trajectories
for a given $\varepsilon$ and $N$. 

Since the ghost propagator has to be calculated for each individual momentum 4-tuple and the number 
of 4-tuples rapidly increases with the lattice size we have restricted ourselves 
to a certain set of momentum 4-tuples for our periodic symmetric lattice.
We have chosen a cylindrical cut similar to Monte Carlo measurements 
such that the individual momentum components $ ap_\mu(k_\mu)$ 
do not differ too strongly from each other. The direction of that 4-vector $ap$ is chosen nearly 
along a lattice 4-volume diagonal.
In a 4-tuple we have chosen the ordered integers $k_\mu$ 
modulo lattice size not to differ by more than one unit and
have assumed a maximal difference $|k_\mu-k_\nu| \le 2$. 

During the Langevin evolution we treated  all 4-tuples leading to non-identical 
measurements as independent ones. This led in our particular realization 
to $27*(N/2)$ 
4-tuples~\footnote{The described choice corresponds to the Leipzig code, in the Parma
code not all of the highest momenta have been measured.}
for which we have measured the propagator.
After obtaining the Langevin time averages we further averaged the propagator 
over 4-tuples which were
equivalent due to lattice symmetries and ended up with the following set of 
inequivalent momentum 4-tuples 
for our lattices chosen with even site numbers 
used in the further analysis:
\begin{eqnarray}
  \label{eq:tuples}
  &&(k,k,k,k) ,\ (k,k,k,k-1),\ (k, k,k- 1,k-1),\ 
  \nonumber \\
  && (k,k-1,k-1,k-1), \quad  k=1,\dots, \frac{N}{2}\,; 
  \\
  &&(k,k,k-1,k-2),\ (k,k-1,k-1,k-2),\ 
 \nonumber \\
 && (k,k-1,k-2,k-2) , \quad   k=2,\dots, \frac{N}{2}\,.
  \nonumber
\end{eqnarray}
Choosing as example $N=8$, out of originally chosen 108 4-tuples 
we get 25 inequivalent 4-tuples for which the propagator has been measured.
A final average has been performed on the results from different trajectories.

In Tables~\ref{tab:statistics1} and  \ref{tab:statistics2}
\begin{table}[!htb]
\begin{center}
\begin{tabular}{|c|cc|cc|cc|cc|cc|cc|}
\hline
 \multicolumn{1}{|c|}{}  
  & \multicolumn{2}{|c|}{$N=8$}
  & \multicolumn{2}{|c|}{$N=10$}
  & \multicolumn{2}{|c|}{$N=12$}
  & \multicolumn{2}{|c|}{$N=14$}
  & \multicolumn{2}{|c|}{$N=16$}
  & \multicolumn{2}{|c|}{$N=20$}
\\
\hline
    \multicolumn{1}{|c|}{}  
  & \multicolumn{1}{|c }{1}
  & \multicolumn{1}{ c|}{2}
  & \multicolumn{1}{|c }{1}
  & \multicolumn{1}{ c|}{2}
  & \multicolumn{1}{|c }{1}
  & \multicolumn{1}{ c|}{2}
  & \multicolumn{1}{|c }{1}
  & \multicolumn{1}{ c|}{2}
  & \multicolumn{1}{|c }{1}
  & \multicolumn{1}{ c|}{2}
  & \multicolumn{1}{|c }{1}
  & \multicolumn{1}{ c|}{2}
\\
 \multicolumn{1}{|c|}{ \raisebox{1.5ex}[-1.5ex]{$\varepsilon$} } 
  & \multicolumn{2}{|c|}{loop}
  & \multicolumn{2}{|c|}{loop}
  & \multicolumn{2}{|c|}{loop}
  & \multicolumn{2}{|c|}{loop}
  & \multicolumn{2}{|c|}{loop}
  & \multicolumn{2}{|c|}{loop}
\\
\hline
    0.010 & 2500&2000  & 1800&1000 & 1200&864 & 1250&822 & 800&384  &   57&57 \\
    0.020 & 1750&1500  & 1250&1000 &  900&850 &  750&650 & 600&241  &   43&43 \\
    0.030 & 1100&1000  &  900&750  &  750&778 &  650&573 & 373&262  &   20&20 \\
    0.050 & 1100&1000  &  900&900  &  750&700 &  650&650 & 284&259  &   39&39 \\
    0.070 & 1100&1000  &  850&750  &  700&770 &  650&591 & 259&192  &   40&40 \\
\hline
\end{tabular}
\end{center}
\caption{Number of one-loop and two-loop ghost propagator measurements for individual
momentum 4-tuples (\ref{eq:tuples}) at different lattices sizes done with the
Leipzig NSPT code at Leipzig PC clusters.}
\label{tab:statistics1}
\end{table}
\begin{table}[!htb]
\begin{center}
\begin{tabular}{|c|c|c|c|c|}
\hline
  $\varepsilon$ & $N=12$    & $N=14$    & $N=16$  & $N=20$   \\ 
\hline
    0.010, all loops & 990 & 676 & 1214 &  735  \\
    0.015, all loops &   0 &   0 & 1025 &  614  \\
    0.020, all loops & 753 & 620 &  968 &  565  \\
    0.030, all loops & 508 & 438 &    0 &    0  \\
    0.040, all loops &   0 &   0 &  539 &  369  \\
    0.050, all loops & 465 &   0 &    0 &    0  \\
    0.070, all loops & 442 & 301 &    0 &    0  \\
\hline
\end{tabular}
\end{center}
\caption{Number of one-, two- and three-loop ghost propagator measurements for 
individual momentum 4-tuples (\ref{eq:tuples}) at different lattices sizes 
performed by the Parma code at Parma and Trento.}
\label{tab:statistics2}
\end{table}
we present our collected statistics for different Langevin steps and lattice volumes
from the Leipzig and Parma code, respectively.
Note that the Leipzig data contain only measurements up to two-loops ($l_{\rm max}=4$).
Combining the measurements from all sources we have rather accurate data of 
the perturbative propagator at each momentum 4-tuple.

The behavior of the different orders of the dressing function $\hat{J}$ is 
shown in Figure~\ref{fig:1} 
for loop numbers $n=1,2,3$ and the
non-loop contribution $n=3/2$
vs. $ a^2\hat p^2$ at different volumes and fixed Langevin step size $\varepsilon=0.01$:
\begin{figure}[!htb]
  \begin{center}
     \begin{tabular}{cc}
        \includegraphics[scale=0.65,clip=true]
         {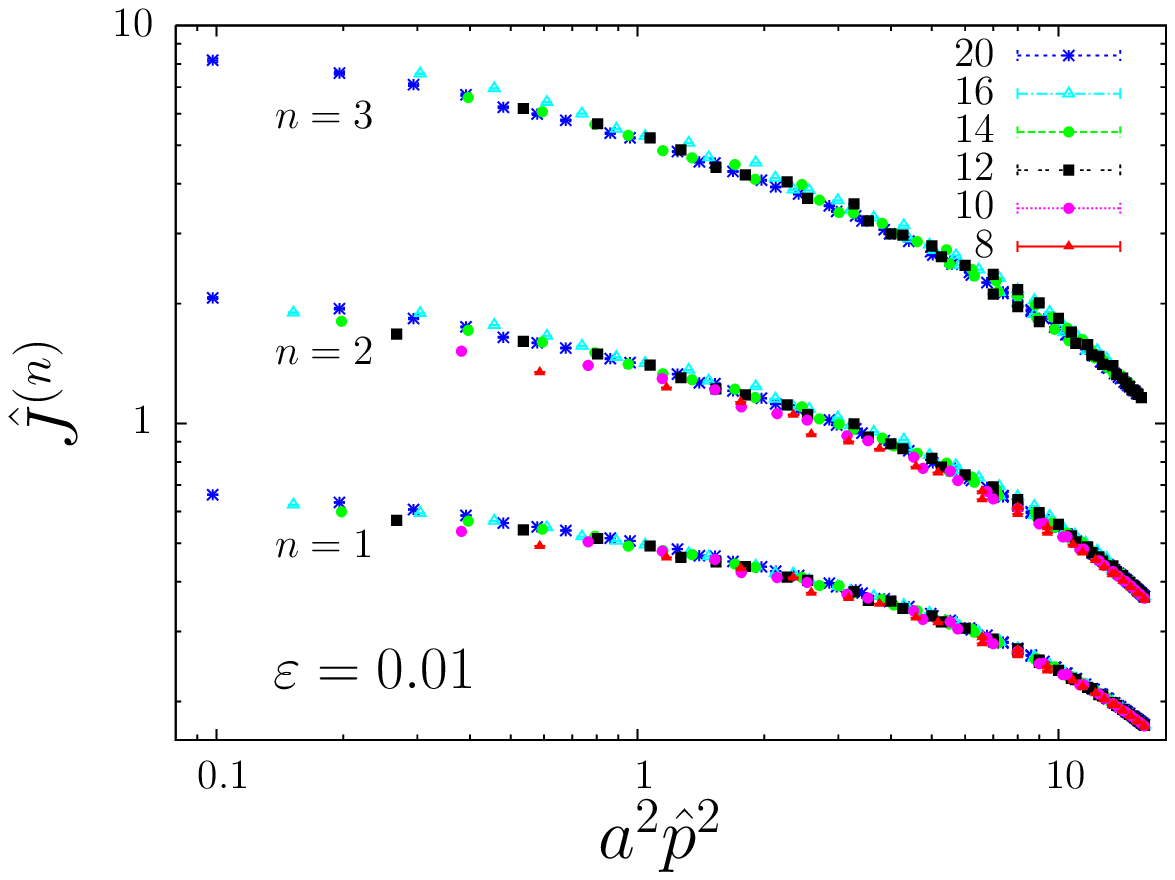}
        &
        \includegraphics[scale=0.65,clip=true]
         {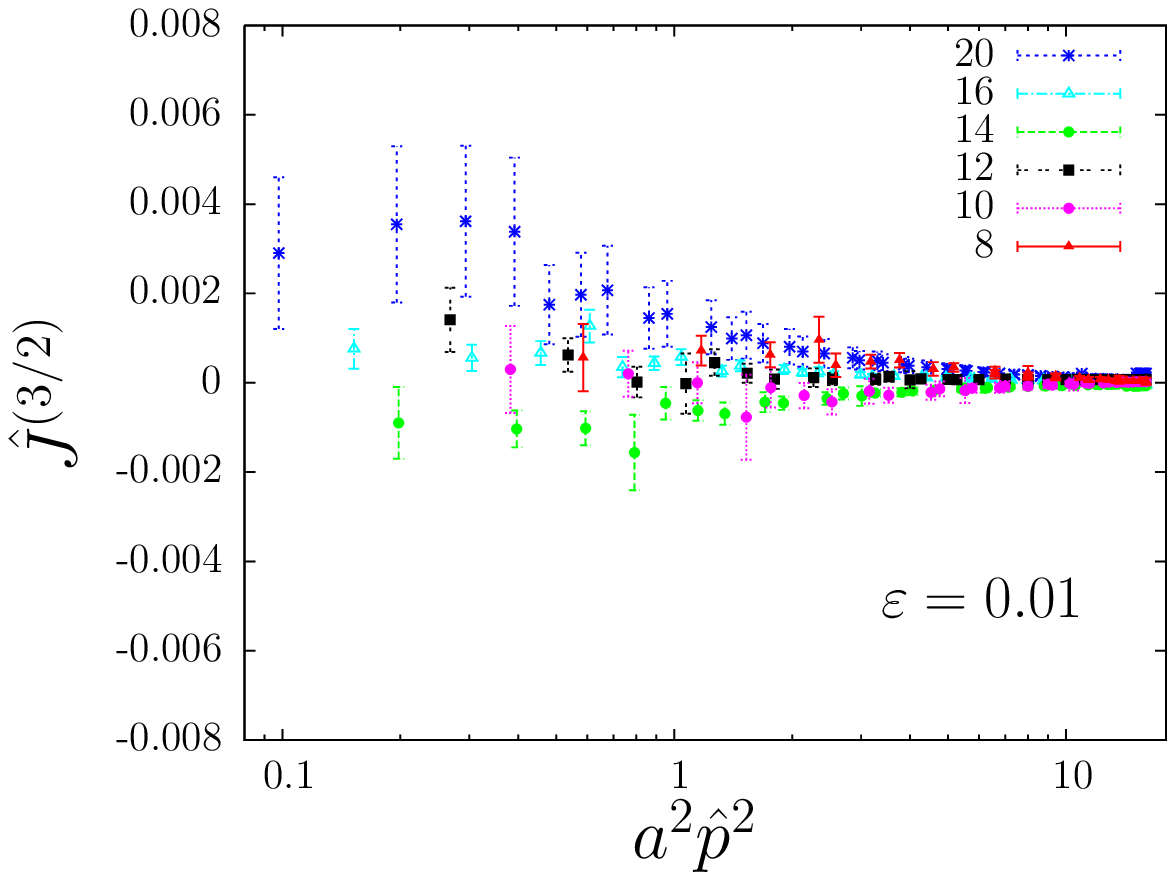}
     \end{tabular}
  \end{center}
  \caption{{Measured ghost dressing function $\hat{J}$ vs. $ a^2\hat p^2$
           for all inequivalent lattice momentum  4-tuples near diagonal for
           $N=8,10,12,14,16,20$ and $\varepsilon=0.01$.
           Left: The one-loop ($\beta^{-1}$), two-loop ($\beta^{-2}$) and three-loop
           ($\beta^{-3}$) contributions,
           right: the vanishing ($\propto \beta^{-3/2}$) contribution.}}
  \label{fig:1}
\end{figure}
We note that the dressing function for the power $\beta^{-3/2}$, 
which is not a loop contribution~\footnote{By construction of the 
FP operator the contribution $\beta^{-1/2}$ is zero in NSPT.},
vanishes to a good accuracy for each individual lattice momentum 4-tuple after 
averaging over the configurations.
This vanishing has to happen already at non-zero $\varepsilon$.
In general, the behavior in the infrared (small lattice momenta) is a bit more noisy. 
For a finite number of measurements the eventually still non-zero non-loop 
contribution $n+\frac{1}{2}$
for some given 4-tuple is by several orders of magnitudes smaller compared 
to the ``neighboring'' $n$-th and $(n+1)$-th loop contributions.

Now we are in the position to perform the extrapolation
to zero Langevin step for the ghost dressing function at each individual 
momentum 4-tuple $(k_1,k_2,k_3,k_4)$ from the available finite $\varepsilon$ measurements
at fixed lattice size.
Since the Langevin system is solved in the Euler scheme in which pieces of $(\varepsilon^2)$
are neglected, we fit the extrapolation of each order of the ghost dressing function $\hat J^{(n)}$
to zero Langevin step by an ansatz of the form $ \hat J^{(n)}(\varepsilon)=
\hat J^{(n)}(p,L) + a \varepsilon + b \varepsilon^2$.
The fit result is a function of the lattice momentum 4-tuple and depends on 
the lattice size.
For most of the cases we used 5 different $\varepsilon$ values, for the largest volume we had
7 Langevin steps for one- and two-loop and 4 such steps for the three-loop extrapolation.

\begin{figure}[!htb]
  \begin{center}
     \begin{tabular}{ccc}
        \includegraphics[scale=0.40,clip=true]
         {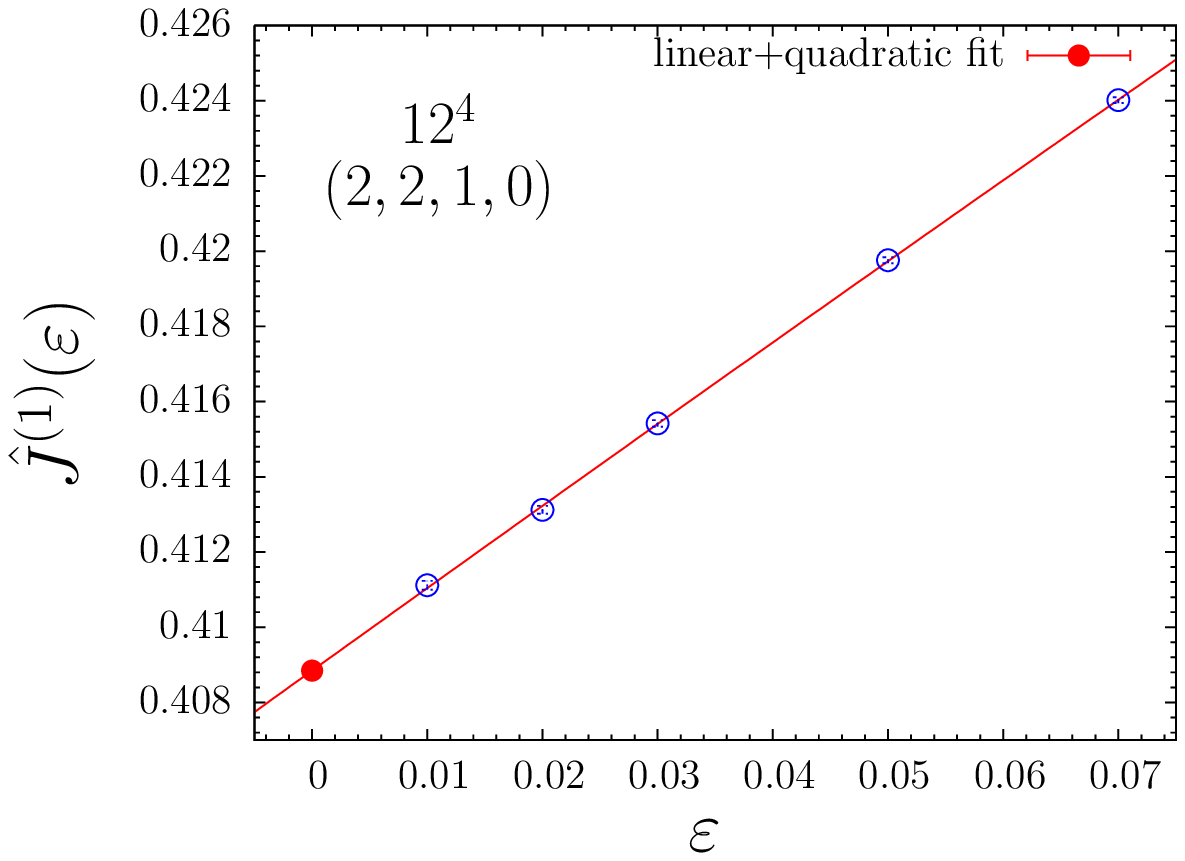}
        &
        \includegraphics[scale=0.40,clip=true]
         {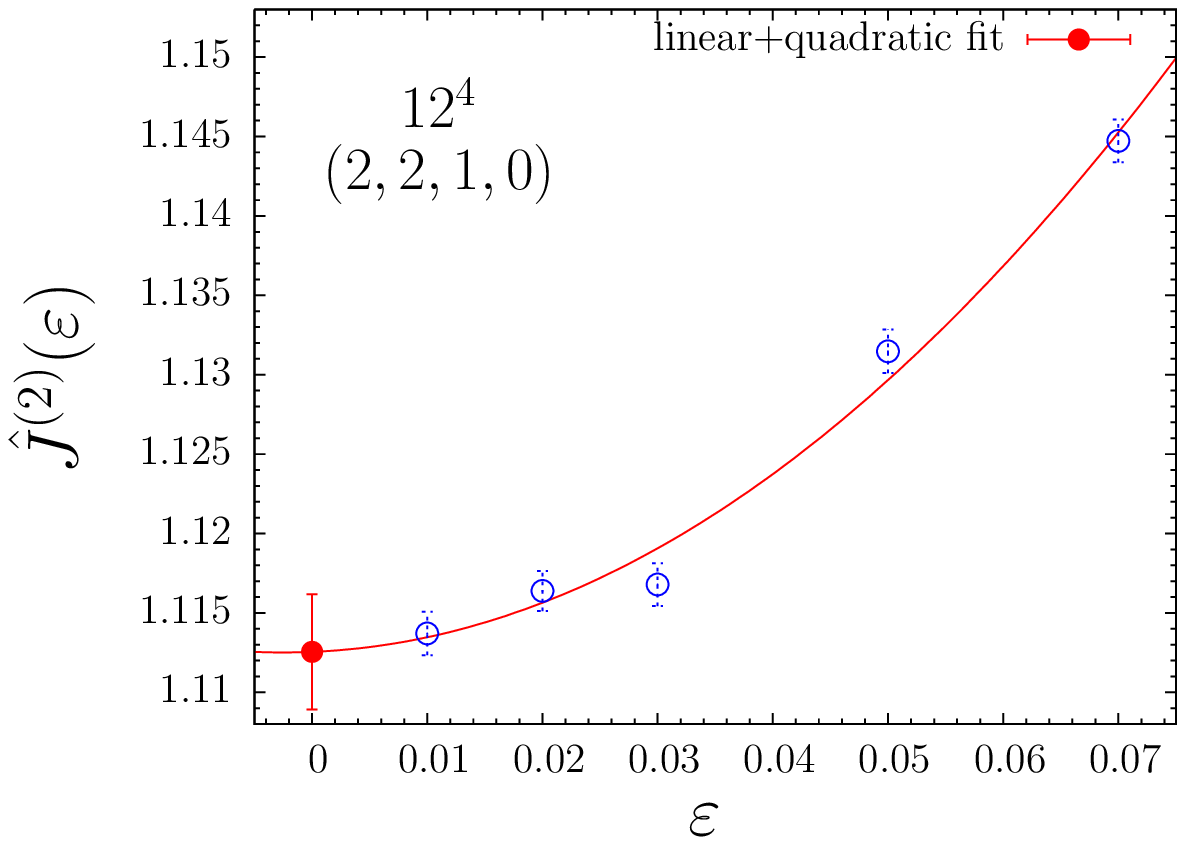}
        &
        \includegraphics[scale=0.40,clip=true]
         {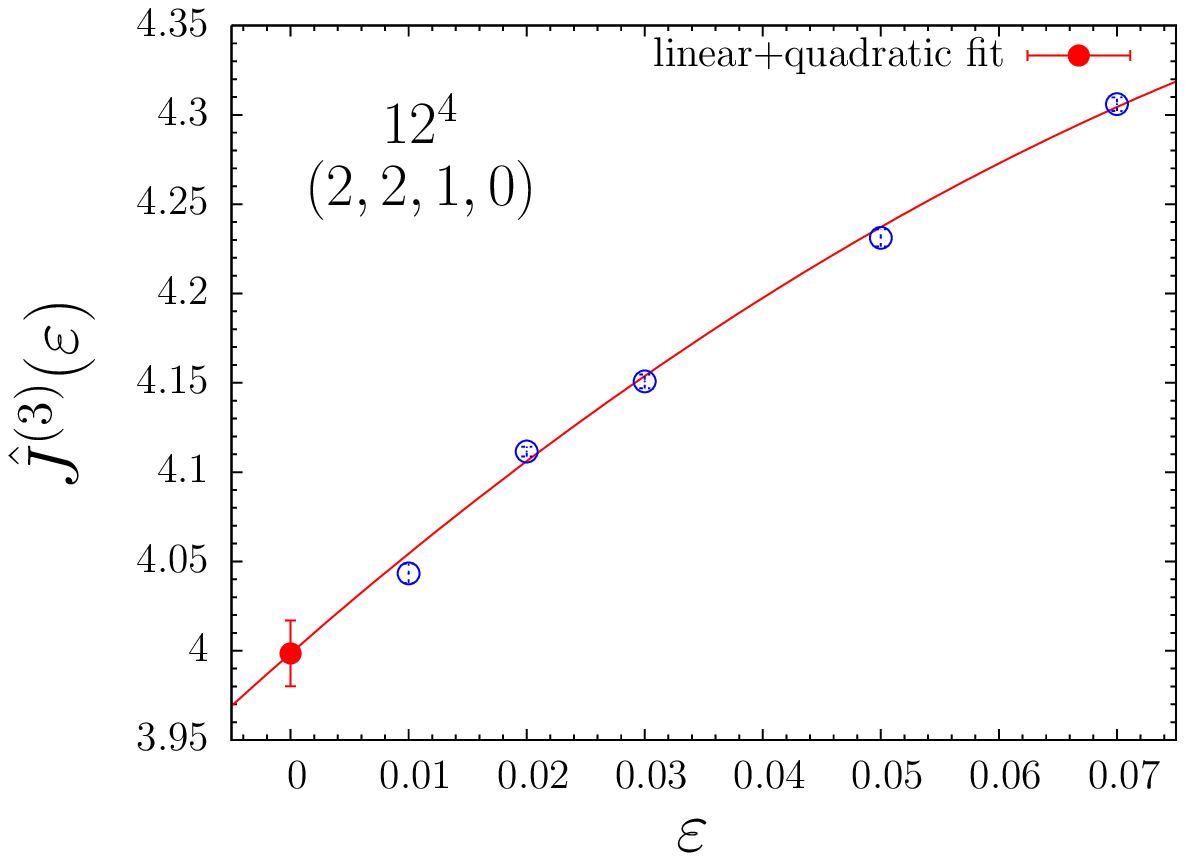}
     \end{tabular}
  \end{center}
  \caption{Extrapolation to $\varepsilon=0$
           of one-, two- and three-loop ghost dressing function for
           lattice size $12^4$ and momentum 4-tuple  $(2,2,1,0)$: the fitting function
           contains both a linear and a quadratic term in $\varepsilon$.}
  \label{fig:2}
\end{figure}

\begin{figure}[!htb]
  \begin{center}
     \begin{tabular}{ccc}
        \includegraphics[scale=0.40,clip=true]
         {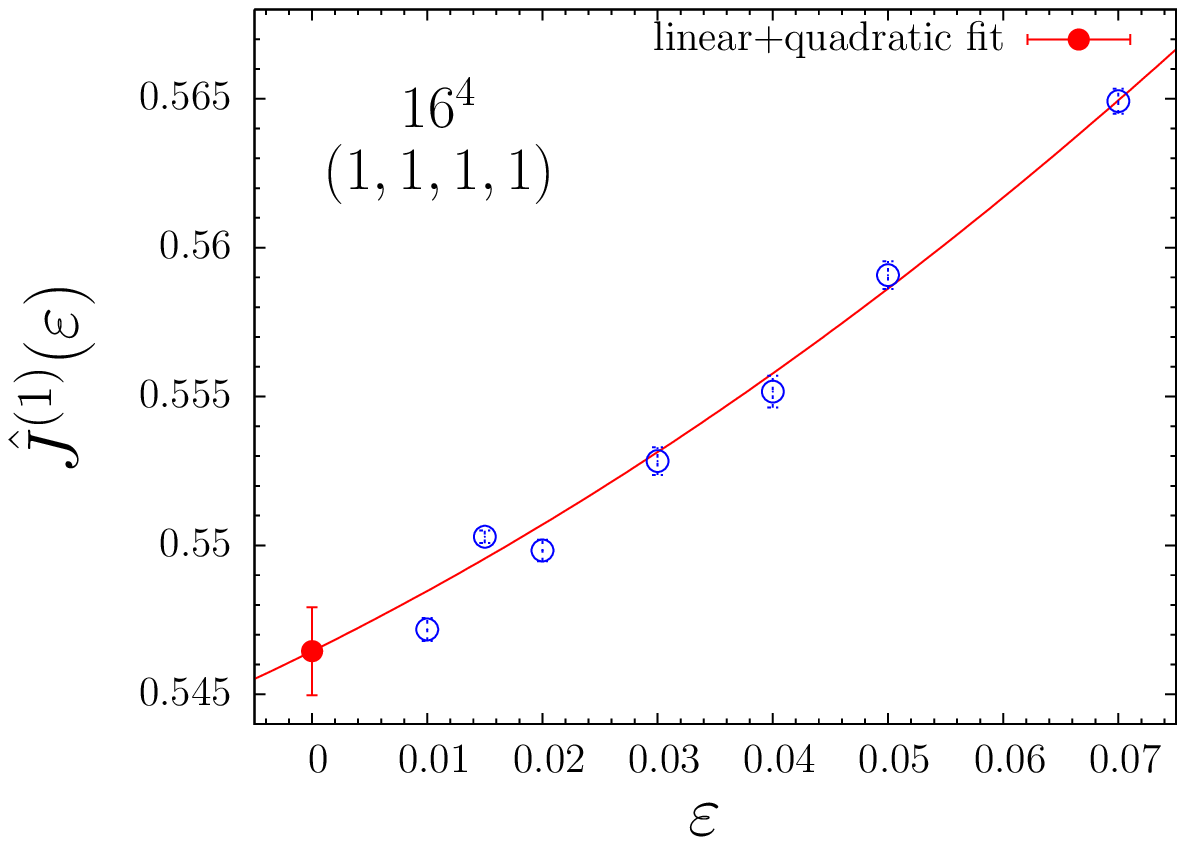}
        &
        \includegraphics[scale=0.40,clip=true]
         {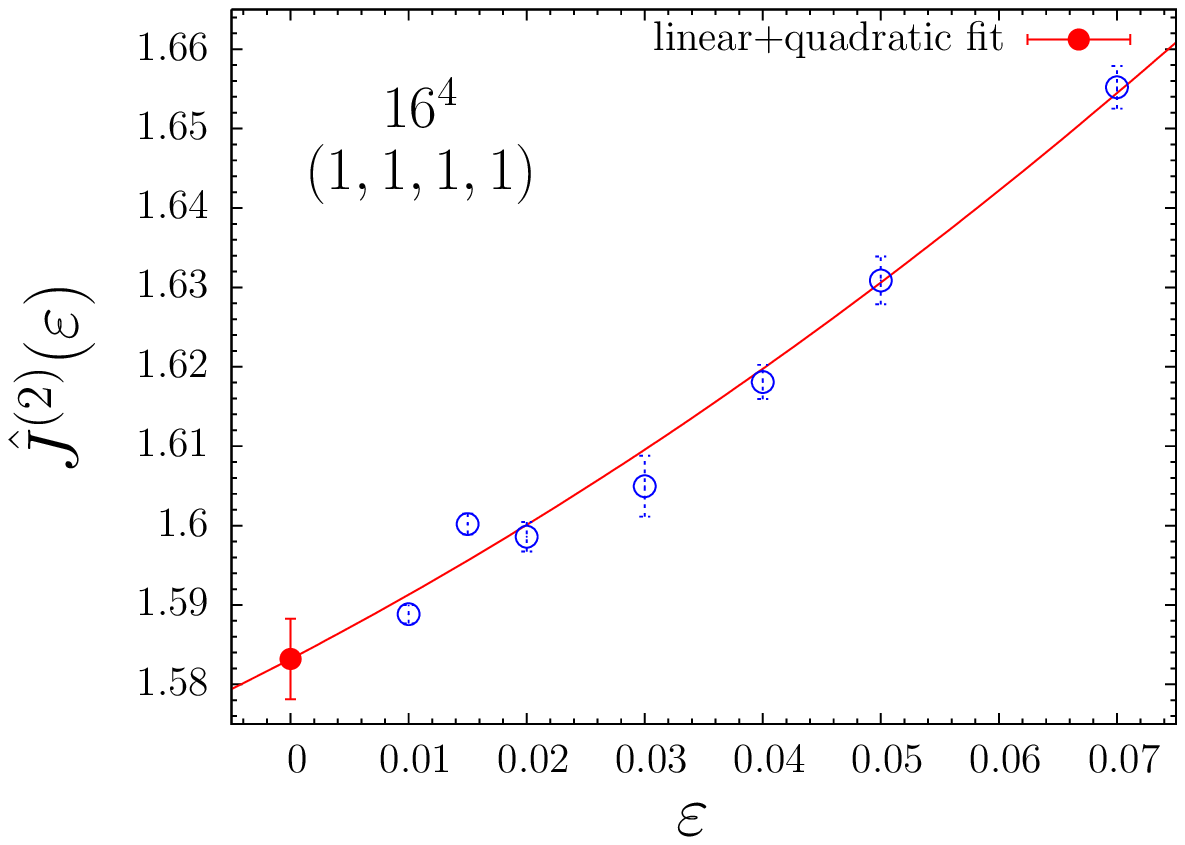}
        &
        \includegraphics[scale=0.40,clip=true]
         {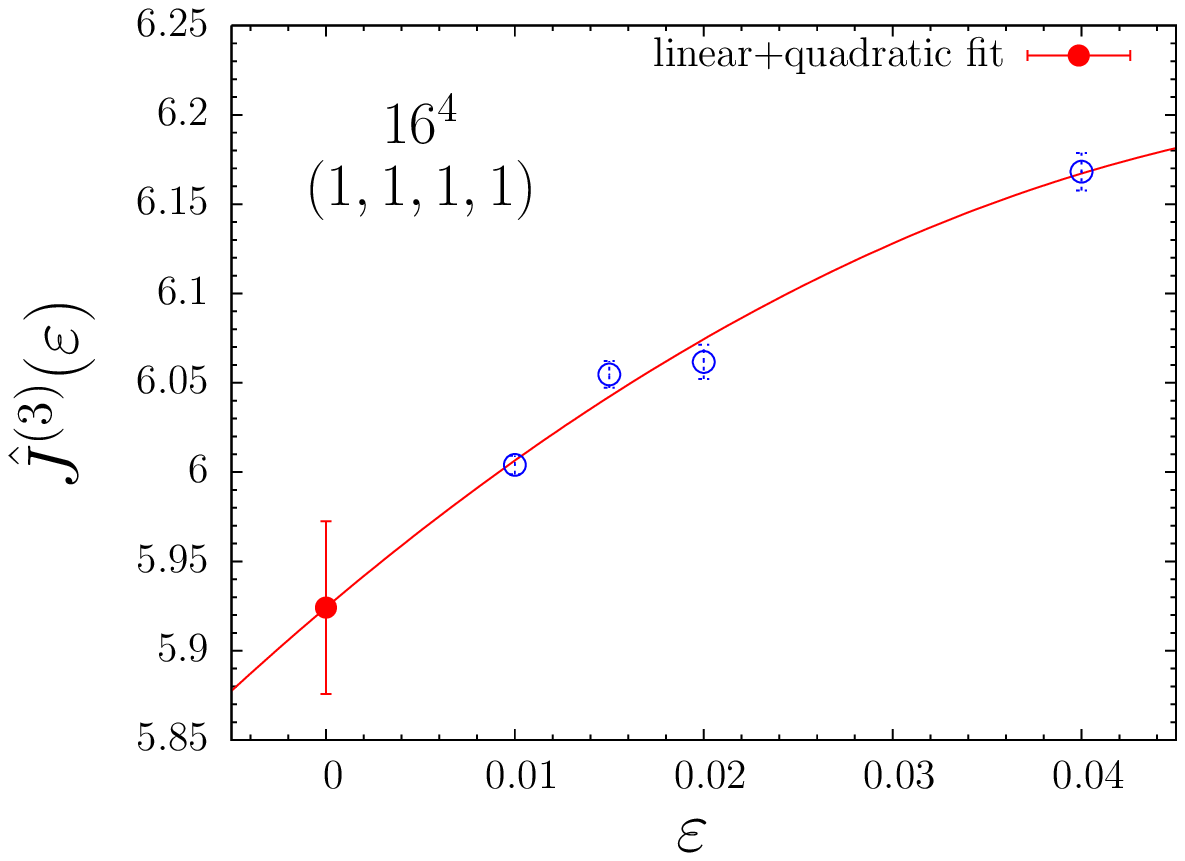}
     \end{tabular}
  \end{center}
  \caption{Same as in Figure~\ref{fig:2} for lattice size $16^4$ 
           and momentum 4-tuple $(1,1,1,1)$.}
  \label{fig:3}
\end{figure}

\begin{figure}[!htb]
  \begin{center}
    \begin{tabular}{cc}
       \includegraphics[scale=0.63,clip=true]{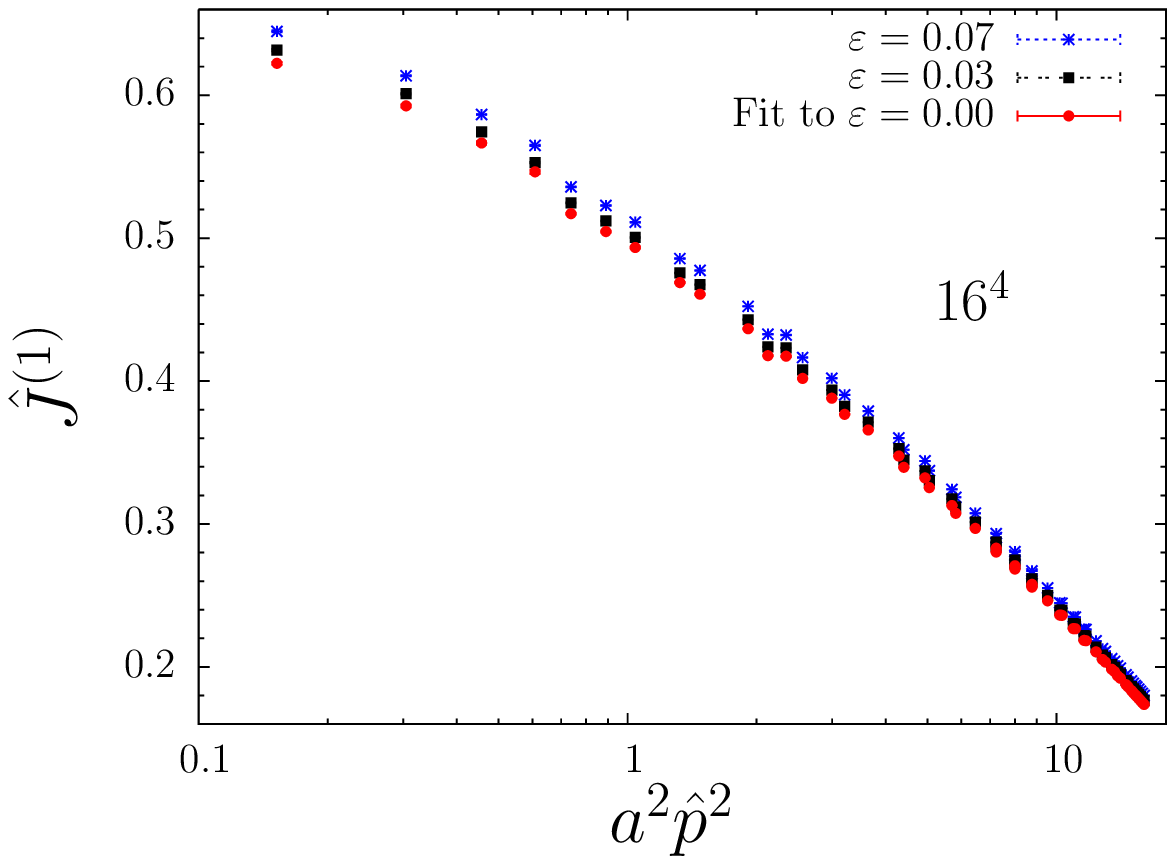}
&
       \includegraphics[scale=0.63,clip=true]{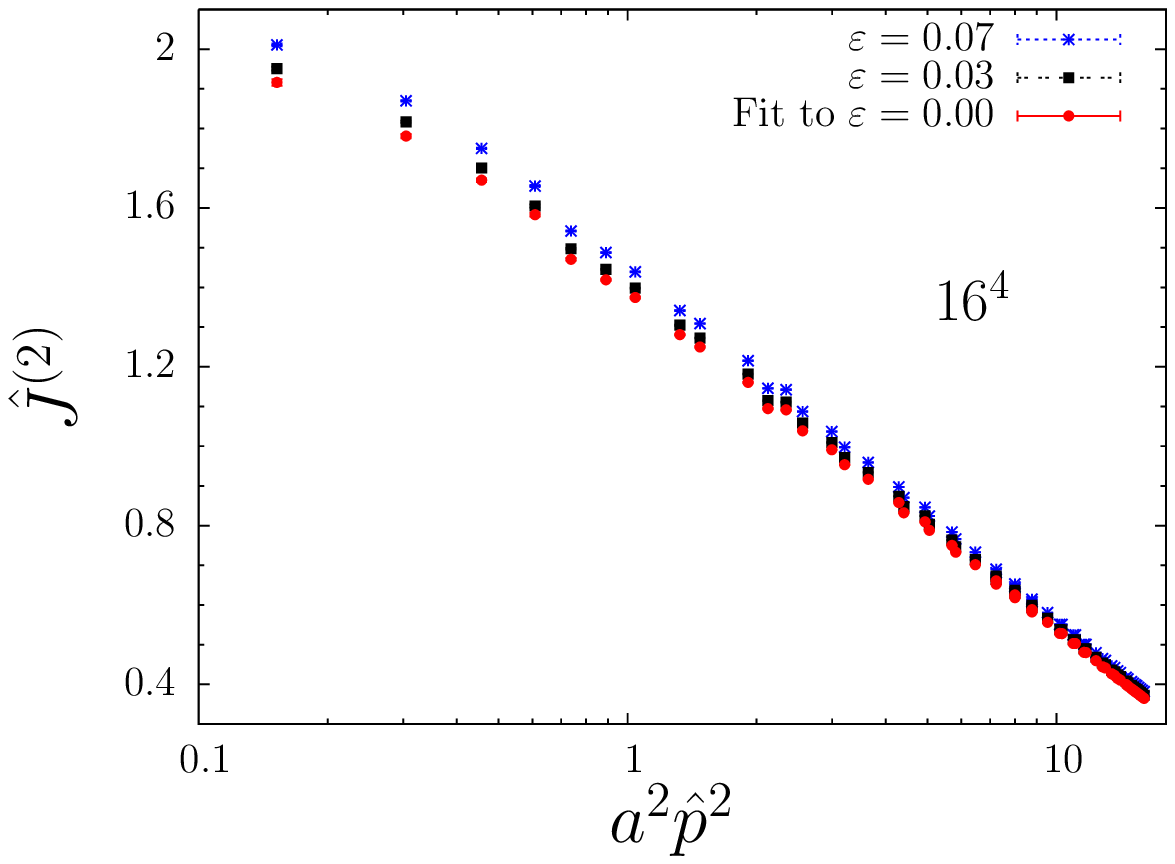}
    \end{tabular}
  \end{center}
  \caption{Extrapolation to zero Langevin step as a function of $a^2 \hat p^2$ for
   a lattice of size $16^4$; left: one-loop dressing function, right: two-loop
   dressing function.}
  \label{fig:extrapolall}
\end{figure}

The extrapolation to zero Langevin step is shown for momentum
4-tuples $(2,2,1,0)$ at $N=12$ and $(1,1,1,1)$ at $N=16$ in Figures~\ref{fig:2}
and~\ref{fig:3}, respectively.
In Figures~\ref{fig:extrapolall} we show the extrapolation of the one- and two-loop 
contributions to zero Langevin step for $N=16$. 

\begin{figure}[!htb]
  \begin{center}
    \begin{tabular}{cc}
       \includegraphics[scale=0.63,clip=true]{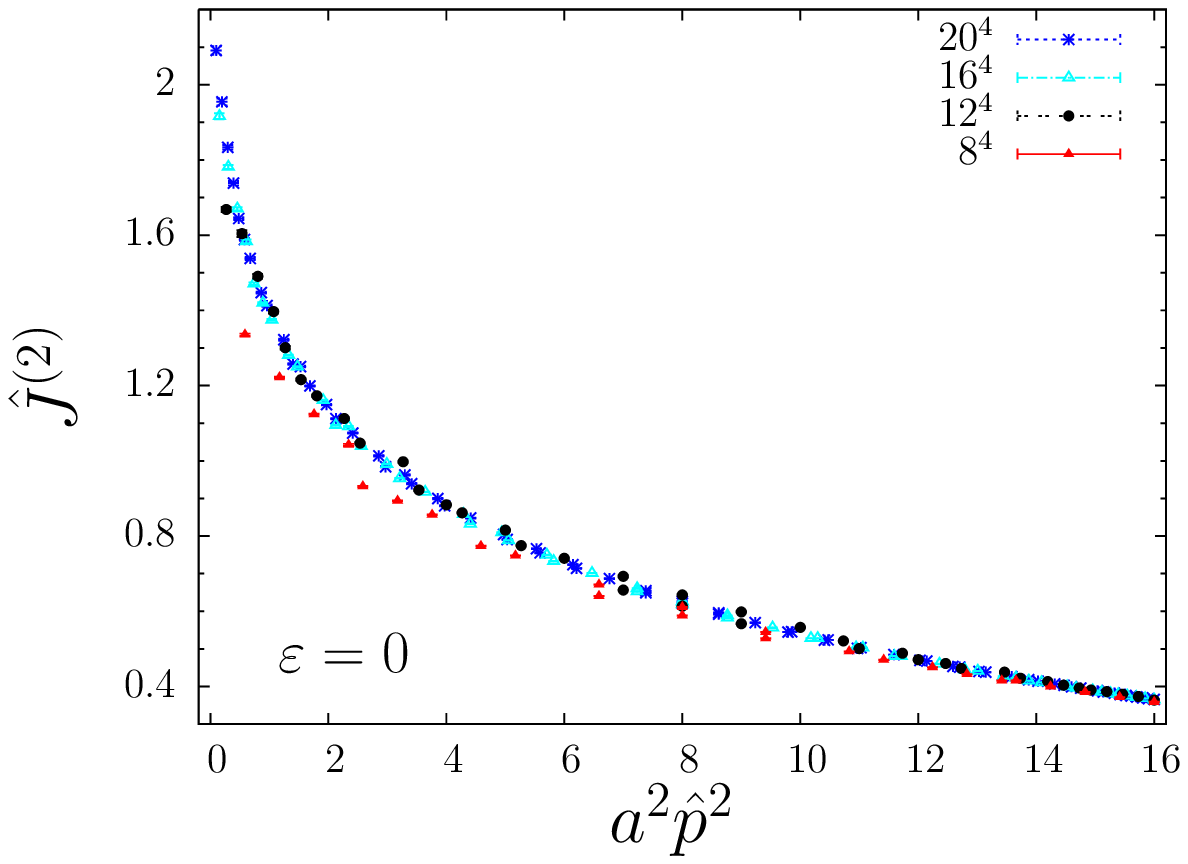}
&
       \includegraphics[scale=0.63,clip=true]{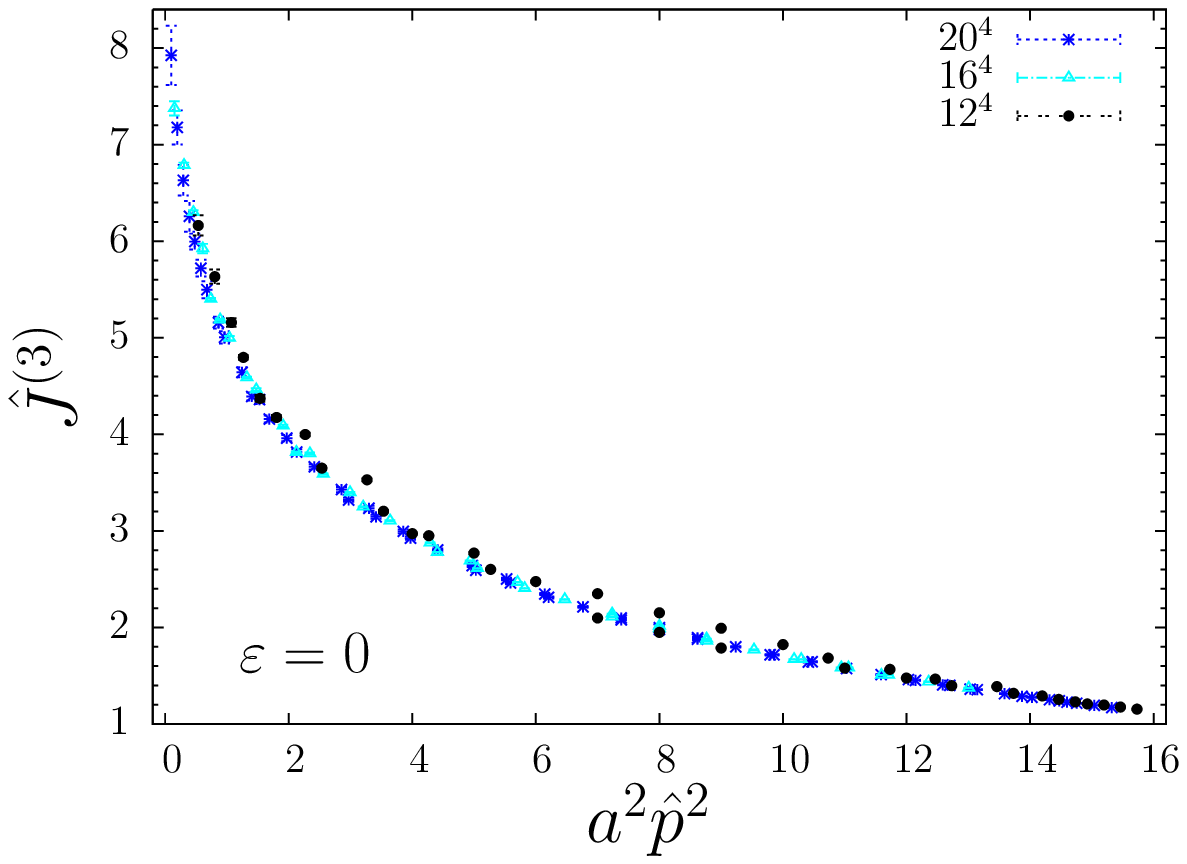}
    \end{tabular}
  \end{center}
  \caption{Volume dependence at zero Langevin step as a function of $a^2 \hat p^2$ for
   different lattice sizes; left: two-loop dressing function, right: three-loop
   dressing function.}
  \label{fig:vol}
\end{figure}

The remaining volume dependence of individual loops contributions vs. $a^2\hat p^2$
at zero Langevin step 
is shown in Figure~\ref{fig:vol} for two-loop and three-loop dressing functions using
some of the lattice volumes. 

One of the aims of this project is to provide referential perturbative 
results for finite lattices and for lattice momenta $(k_1,k_2,k_3,k_4)$ 
without any restriction. For this purpose, no further extrapolation is 
necessary. An exhaustive comparison
with Monte Carlo results is not attempted in this paper. 

To reproduce LPT results in the infinite-volume limit at vanishing 
lattice spacing, however, different lattice sizes have to be studied 
in parallel as explained in the following.

\subsection{Fitting the data}
\label{subsec:fitting}

Our aim is to extract the finite constants $J_{i,0}$ for loop order $i$
in the perturbative expansion of the 
lattice ghost dressing function (\ref{Zghostbeta3loop}) 
in infinite volume and in the continuum limit $a \rightarrow 0$
as accurate as possible.
As pointed out in section \ref{sec:standard-LPT}, at any loop, a computation 
would yield a sum of a finite contribution plus logarithmic terms. 
To be definite, we recall that at 
one-loop order one gets
\begin{equation}
  J^{(1)}(a,p)= J_{1,1} \log (pa)^2 + J_{1,0} \, ,
  \label{eq:J1a0L0}
\end{equation}
with the known coefficient $J_{1,1}$ given in (\ref{J1loop}) and similar expressions
for higher loops.
In general, at loop order $i$ one gets terms up to power $\left(\log(pa)^2 \right)^i$, 
and all logarithmic contributions arise from anomalous dimensions and the beta function.
Given what is known in the RI'-MOM scheme, in our present computation we can aim at three loops. 
In particular, reproducing the known one-loop lattice constant $ J_{1,0}$ 
is an obvious check for our NSPT results. 
However, any NSPT computation is performed at both finite lattice spacing and finite lattice size. 

Let us first of all discuss the influence of the finite lattice spacing. 
Any lattice computation does not have any explicit reference to the lattice spacing $a$; in a 
non-perturbative computation this is always fixed a posteriori by matching to a 
physical scale. In the NSPT propagator computations one always handles dimensionless quantities 
which are fixed once a 4-tuple $(k_1,k_2,k_3,k_4)$ (with integer $k_\mu$) is selected.
One has no direct access to a (dimensionful) momentum
$p_{\mu}$ (see (\ref{eq:plainMOM})),
but only handles (notice the explicit reference to lattice size $L=Na$, with 
$N$ a pure number)
\begin{equation}
  p_{\mu}a = \frac{2 \pi k_{\mu}}{N} \,.
  \label{eq:identity}
\end{equation}
While in the $p_\mu$ definition the lattice size enters explicitly, one is 
left with the combination $pa$ as variable, which for $a \neq 0$ is the signature
of finite lattice spacing also in infinite volume. 
Taking into account $a \neq 0$, {\em e.g.} at one loop, 
the dressing function results in 
\begin{equation}
  J^{(1)}(pa)= J_{1,1} \log (pa)^2 + J_{1,0} (pa)\ ,
  \label{eq:J1aL0}
\end{equation}
instead of (\ref{eq:J1a0L0}).
The non-logarithmic constants $J_{i,0}(pa)$ for loop number $i$
are functions which encode the lattice artifacts. Since we expect these functions to be compliant 
to lattice symmetries, there is an obvious way to estimate the $pa \rightarrow 0$ limit 
by means of a {\em hypercubic-invariant Taylor series}~\cite{Di Renzo:2006wd}. 
We collect measurements for (moderately) small values of $pa$ and fit our results 
for the non-logarithmic part of a generic i-th loop to a formula 
which is fixed by the physical dimension and the scalar nature of the observable at hand
\begin{equation}
  J_{i,0}(pa) =  J_{i,0} 
  + c_{i,1} \, (pa)^2 
  + c_{i,2} \, \frac{(pa)^4}{(pa)^2} 
  + c_{i,3} \, (pa)^4
  + c_{i,4} \, \left((pa)^2\right)^2
  + c_{i,5} \, \frac{(pa)^6}{(pa)^2} 
  + \cdots \,,
  \label{eq:H4Taylor}
\end{equation}
in which $(pa)^n$ stands for the hypercubic invariant $\sum_{\mu}(p_{\mu}a)^n$. 
This is a fit with 6 parameters
in which no (possible) subleading logarithms (proportional to $O(a^2)$)
are taken into 
account~\footnote{This is quite a strong hypothesis; 
the effectiveness of such a procedure can only validated a posteriori
({\em e.g.} by inspecting $\chi^2$ values). 
In infinite volume these logs are calculable in principle. For the
quark propagator and quark bilinears this has been done in~\cite{Constantinou:2009tr}.
}. 
Apart from neglecting those logarithms, (\ref{eq:H4Taylor}) is a systematic expansion in  
powers of $pa$. Going to next order, 
would require the inclusion of $5$ extra parameters. 
Since 
$J_{i,0}$ ({\em i.e.} the value at $pa=0$) is the final number we are interested in, 
it is worth to stress that in principle the more we approach $pa=0$, the better the fit is 
expected to work, at any fixed accuracy (in our case, $O\left((pa)^6,\dots\right)$). 
On the other side, the higher values of $pa$ we take into account, the more terms 
in the Taylor series are expected to give a numerically significant contribution. 

Now we have to discuss how the finite lattice size has to be taken into account.
In~\cite{Di Renzo:2006wd} it was pointed out that those effects can be large when 
an anomalous dimension is in place. Having at hand a variety of lattice extents, in this study 
we address a careful assessment of these effects (the main ideas entering the procedure can be 
found in~\cite{NSPT_FeynGauge,DiRenzo:2007qf}). 

To take finite L effects into account, we consider as further generalization the ansatz 
\begin{equation}
  J^{(1)}(pa,pL)= J_{1,1} \log (pa)^2 + J_{1,0;L} (pa,pL)\,,
  \label{eq:J1aL}
\end{equation}
using the one-loop contribution as example.
Dimensional arguments suggest a dependence on $pL$. Notice that this is not an 
irrelevant effect: a $pL$-effect would be there also in a continuum computation 
on a finite volume. In this notation $J_{1,0} (pa)$  from (\ref{eq:J1aL0})
corresponds to the infinite lattice size limit
$J_{1,0} (pa) = J_{1,0;L} (pa,\infty)$. 
Finite size effects can now be parametrized as
\begin{eqnarray}
  J^{(1)}(pa,pL) & = & J_{1,1} \log (pa)^2  + J_{1,0} (pa) + 
  \left( J_{1,0;L} (pa,pL) - J_{1,0} (pa) \right) 
  \\
  & \equiv & J_{1,1} \log (pa)^2  + J_{1,0} (pa) + \delta J_{1,0}(pa,pL) 
  \,,
  \nonumber
\end{eqnarray}
which is, however, not yet useful for a fitting procedure.

To make the ansatz usable for a fit purpose we  once again perform the
expansion (\ref{eq:H4Taylor}) of $J_{1,0} (pa)$ in hypercubic invariants 
and take into account the (formal) continuum limit of $\delta J_{1,0}(pa,pL)$ 
\begin{equation}
  \delta J_{1,0}(pL) \equiv \delta J_{1,0}(pa=0,pL) \,.
\end{equation}
Here $pa$ corrections to the $pL$ correction  $\delta J_{1,0}(pL)$ are supposed to be 
{\em corrections on corrections} and will be neglected.
 
For the non-logarithmic part of an i-th loop this procedure results in
\begin{equation}
  \label{eq:aqqL}
   J_{i,0}(pa,pL) =  J_{i,0} (pa) + \delta J_{i,0}(pL) + \; \cdots\ ,  
  \nonumber 
\end{equation}
where $J_{i,0}(pa)$ is given by the expansion (\ref{eq:H4Taylor}).
Now (\ref{eq:aqqL}) is indeed amenable to a fit. 
The key observation is the trivial equality (\ref{eq:identity})
$p_{\mu}L = p_{\mu}aN = 2 \pi k_{\mu}$ 
which can be treated as follows:
Neglecting $pa$ corrections to $\delta J_{i,0}(pL)$ implies 
that measurements 
at the {\em same} fixed 4-tuple $(k_1,k_2,k_3,k_4)$ but at  {\em different} lattice sizes $L/a$  
are affected by the {\em same} $pL$-effect (while $pa$ values are different).

This leads to a practical way of implementing the fit encoded in~(\ref{eq:aqqL})
without assuming a functional form for the $pL$ effects. 
Suppose we want to fit $J_{1,0}(pa,pL)$. 
First we select an interval $[(pa)^2_{\rm min},(pa)^2_{\rm max}]$, say $[0.25,6.25]$ for $(pa)^2$. 
Then we select a collection of lattice sizes $\{N = 8,10,12,14,16,20\}$ 
where measurements are available and identify all those 4-tuples $k_{\mu} = (k_1,k_2,k_3,k_4)$ 
(resulting in values of $(pa)^2$ inside the chosen interval)
which belong to {\rm each} of the lattice sizes $N = L/a$.
In the case at hand, these are 
\begin{equation}
  \{(1,1,1,0),\,(1,1,1,1),\,(2,1,0,0),\,(2,1,1,0),\,(2,1,1,1),\,(2,2,1,0),\,(2,2,1,1)\}\ ,
\end{equation}
for which we choose the  measured ghost dressing functions at one-loop
after subtracting the logarithmic behavior $J_{1,1} \log (pa)^2$ assumed to be known.
For our 7 4-tuples and 6 lattice sizes, we have just selected 42 measurements
or data points. 

This choice so far still does not fix the overall normalization of the data.
For that purpose we notice that finite-volume effects decrease both with increasing 
momentum squared and increasing lattice size.
Therefore, we choose as reference fitting point -- for a good approximation to
$pL=\infty$ -- a (log subtracted) data point from the largest available lattice 
at $(pa)^2$ approximately equal to the largest momentum squared of the chosen data set.
In our example we add the measurement taken on the $N_{\rm max}=20$ 
lattice for the 4-tuple $k_{\mu \;{\rm max}} \equiv (4,4,4,4)$. 
This choice leads to $(pa)^2=6.31655$ compared to the maximal $(pa)^2=6.1685$
from tuple $(2,2,1,1)$ at $N=8$.
All together we now have $43$ data points and $13$ parameters to fit: 
6  parameters $J_{1,0}$ and $\left\{c_{1,l}  \,; l=1,\dots,5\right\}$ and
7 $\delta J_{1,0}(pL(k))$ for each of the 7  4-tuples $k_{\mu} \neq k_{\mu \;{\rm max}}$.
By assumption we have $\delta J_{1,0}(pL_{\rm max}(k_{\rm max}))=0$
for the normalization point. 

The application of the described criteria to select measurements 
is analogous for higher loop cases.
Before fitting the data in a chosen momentum range, all logarithmic pieces (supposed to be known)
have to be subtracted.

Finally let us discuss the effectiveness of our fitting procedure. A list of 
requirements to be fulfilled should be recalled:
\begin{itemize}
\item 
A number of measurements on different $N=L/a$ lattice sizes should be available.
\item
An interval $[(pa)^2_{\rm min},(pa)^2_{\rm max}]$ has to be singled out 
in which a hypercubic Taylor expansion (with a manageable number of 
terms) for the quantity $J_{i,0} (pa)$ has to be effective.
\item
The number of 4-tuples which enter our procedure has to be such that the 
total number of data points is sufficiently large with respect to the number of 
fit parameters. That number has to be not too large.
\item
It is assumed 
that the data point associated to $k_{\rm max}$ is virtually free of finite size effects. 
\item
All the procedure results in a sufficiently stable fit.
\end{itemize}

\begin{figure}[!htb]
  \begin{center}
    \begin{tabular}{cc}
       \includegraphics[scale=0.67,clip=true]{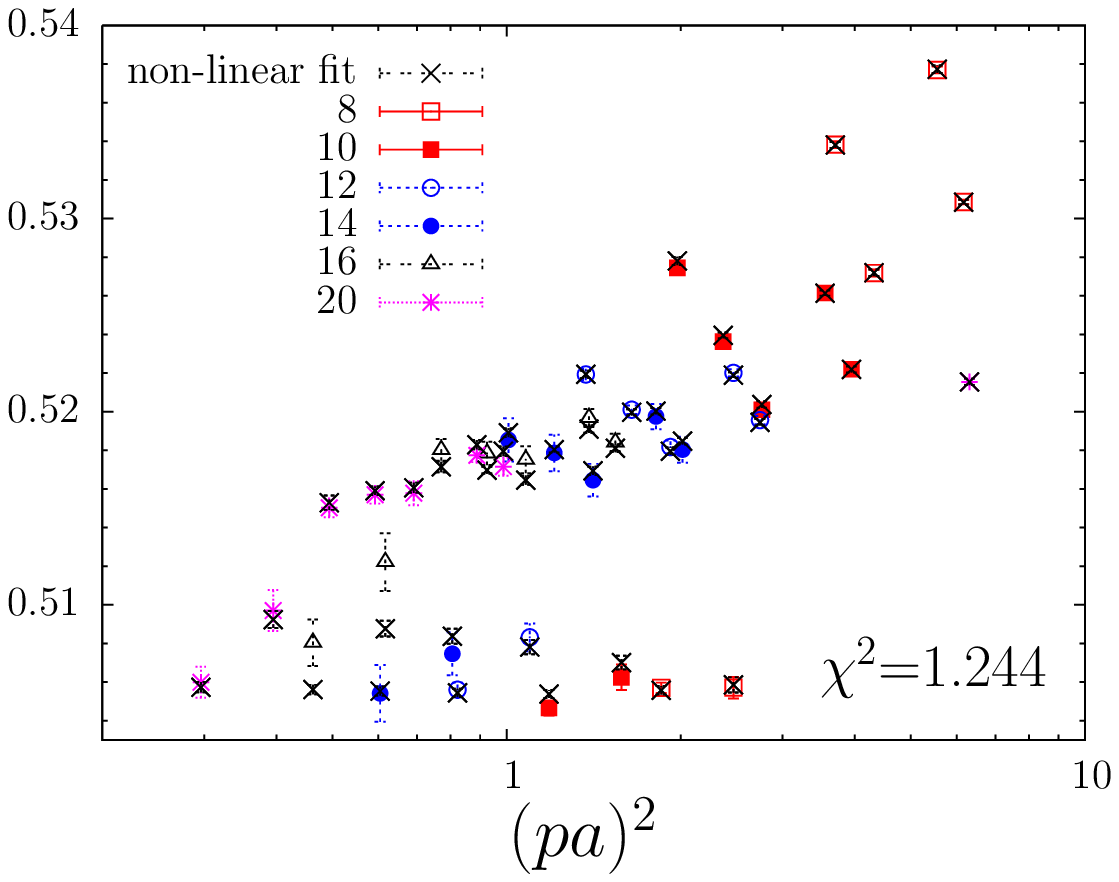}
&
       \includegraphics[scale=0.67,clip=true]{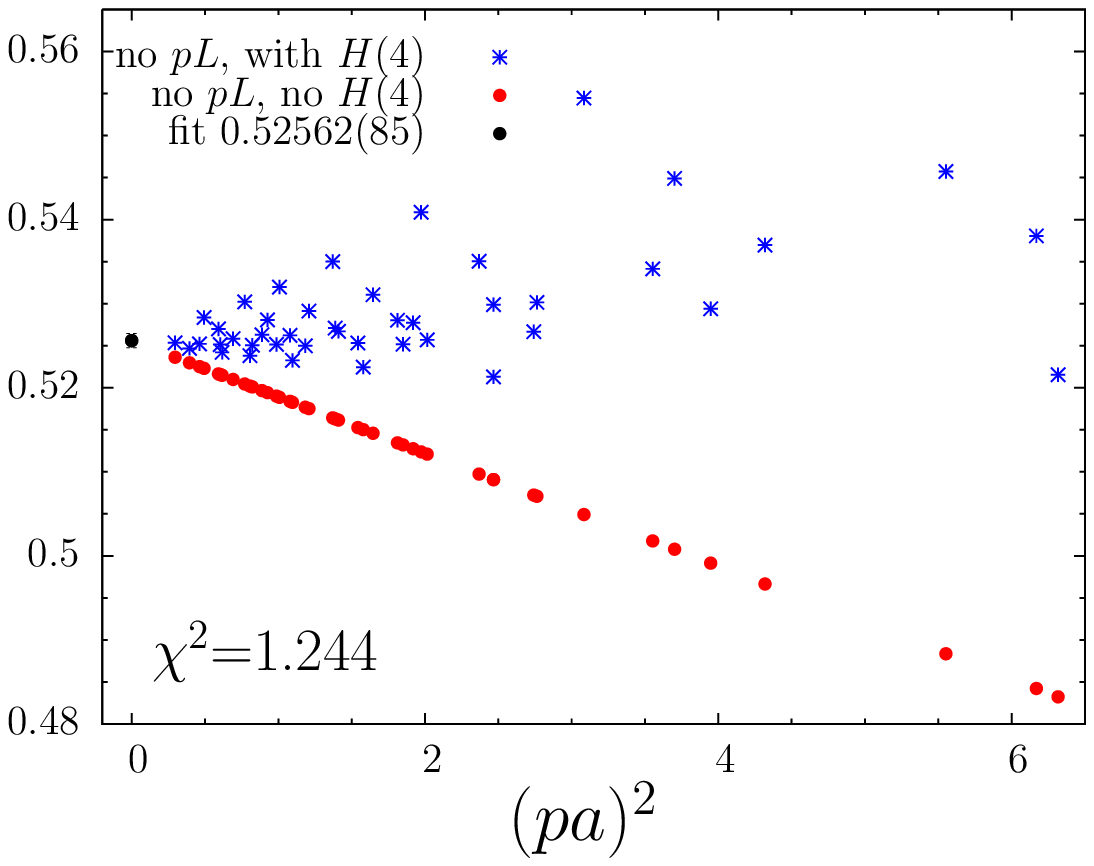}
    \end{tabular}
  \end{center}
  \caption{Fitting of the one-loop non-logarithmic coefficient. 
           Left: Data points after logarithmic subtraction for various lattice sizes $N$
                 compared to points using the non-linear fit (\ref{eq:aqqL}) with 
                 (\ref{eq:H4Taylor}). 
           Right: Stars are data after correction for finite volume effect,
                  i.e. they represent $J_{i,0}(pa)$; full circles are fit points after 
                  correcting both finite volume and some hypercubic effects with the exception of 
                  those proportional to the coefficients $c_{i,1}$ and $c_{i,4}$ in (\ref{eq:aqqL}).}
  \label{fig:pLatWork1}
\end{figure}

\begin{figure}[!htb]
  \begin{center}
    \begin{tabular}{cc}
       \includegraphics[scale=0.67,clip=true]{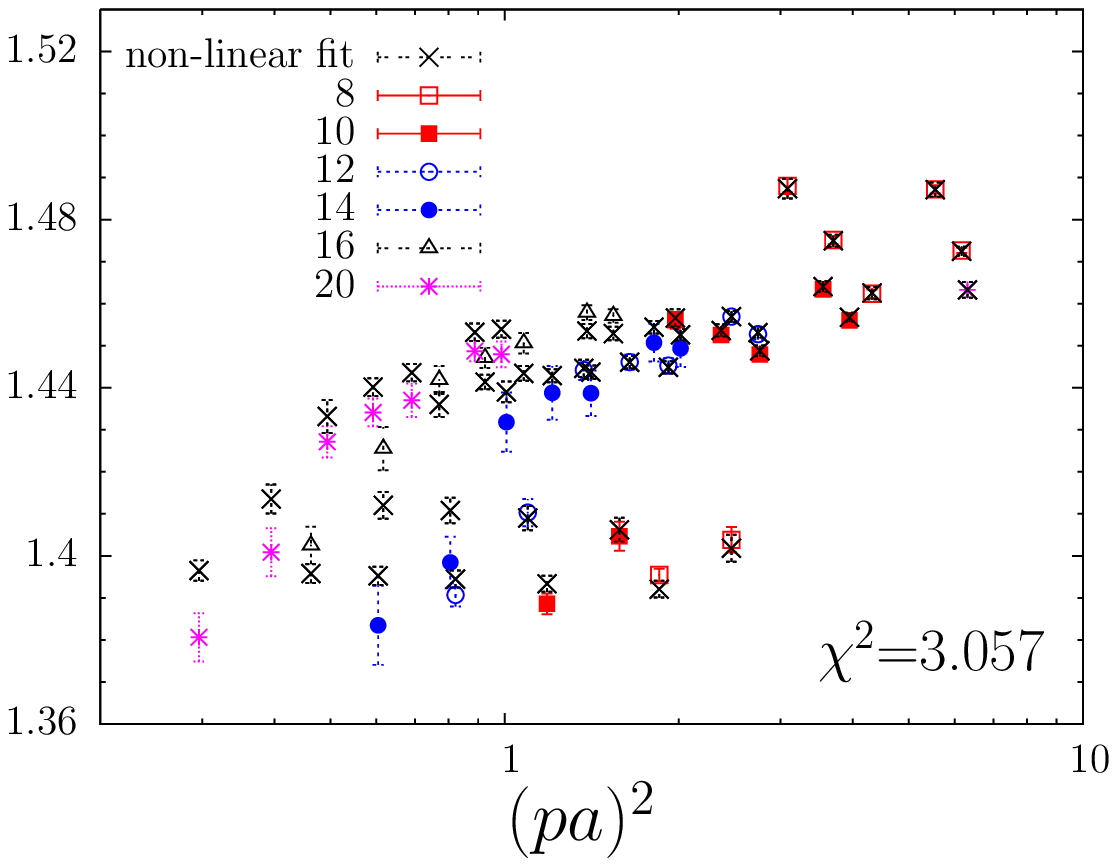}
&
       \includegraphics[scale=0.67,clip=true]{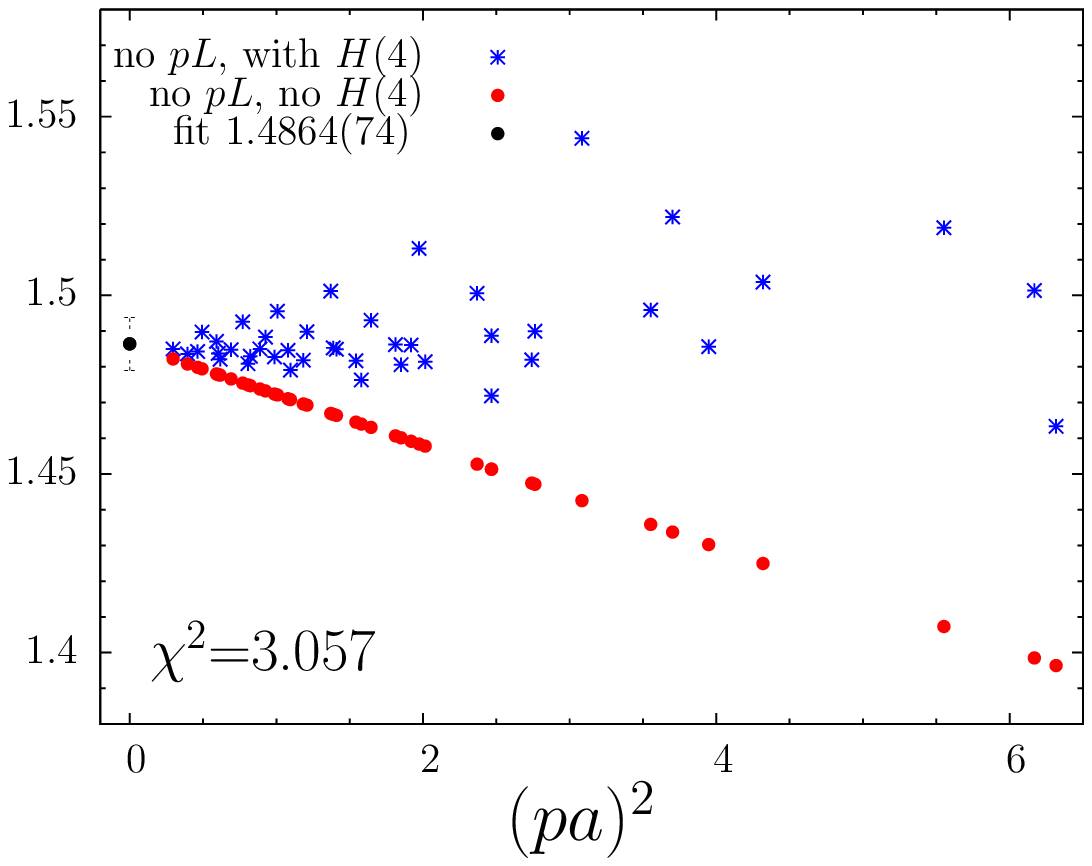}
    \end{tabular}
  \end{center}
  \caption{Same as in Figure~\ref{fig:pLatWork1} for the two-loop non-logarithmic coefficient.} 
  \label{fig:pLatWork2}
\end{figure}

The fulfillment of all these requirements has to be assessed a posteriori. 
Figures~\ref{fig:pLatWork1} and \ref{fig:pLatWork2} shows the method at work at one
loop and two loops.
Looking at the left Figures~\ref{fig:pLatWork1},\ref{fig:pLatWork2} we observe that the numerical data 
(extrapolated to $\varepsilon=0$) from the chosen set of all lattice sizes 
-- with all logarithmic contributions subtracted -- scatter significantly.
Switching off the 
finite volume $O(p L)$ corrections ($\delta J_{i,0}(pL)$ in the fit form~(\ref{eq:aqqL}))
in the right Figures~\ref{fig:pLatWork1},\ref{fig:pLatWork2},
the stars (blue in color online) line up in 'rows' according to
the different hypercubic invariants at infinite lattice volume.
The reference point (here the rightmost point) is of course unchanged compared to its position
in the left Figures.
Finally, after removing also the non-rotational hypercubic 
artefacts (leaving only $c_{i,1} \ne 0$ and $c_{i,4} \ne 0$ in (\ref{eq:aqqL})),
we obtain a smooth (almost linear) curve  formed by the full circles (red in color online) which
directly points towards the fitting constant $J_{i,0}$ 
corresponding to the zero lattice spacing limit.
The quality of the fitting procedure can be estimated by comparing the data with the fit points,
as it can been seen in the left Figures.

\section{Results}
\label{sec:results}

\subsection{Finite constants in LPT up to three loops}

In order to determine the dressing function for the ghost propagator $J(a,p,\beta)$
up to three-loop order on the lattice we have to compute the constants $J_{k,0}$
in the way outlined in Section \ref{subsec:fitting}.
They fix all other coefficients of the subleading logarithms as it can be
seen in formulae (\ref{J1loop}),  (\ref{J2loop}) and (\ref{J3loop}). 
These relations also show that the coefficients of a certain
order in perturbation theory depend on all lower-order results.

One essential point in the fit procedure described in Section \ref{subsec:fitting} is the
existence of a stable fitting window $(pa)^2_{\rm min} \le (pa)^2 \leq  (pa)^2_{\rm max}$. 
Practically,
it is defined by the value of the quadratic error $\chi^2$ of a nonlinear regression fit.
It is not possible to give a
prescription to fix  {\it a priori} an {\it optimal} value for $\chi^2$
besides the general condition $\chi^2 \gtrsim 1$.
It has to be determined for each case (i.e. fit function $J^{(k)}(pa)$) by
inspection.
\begin{figure}[!htb]
  \begin{center}
    \begin{tabular}{cc}
       \includegraphics[scale=0.63,clip=true]{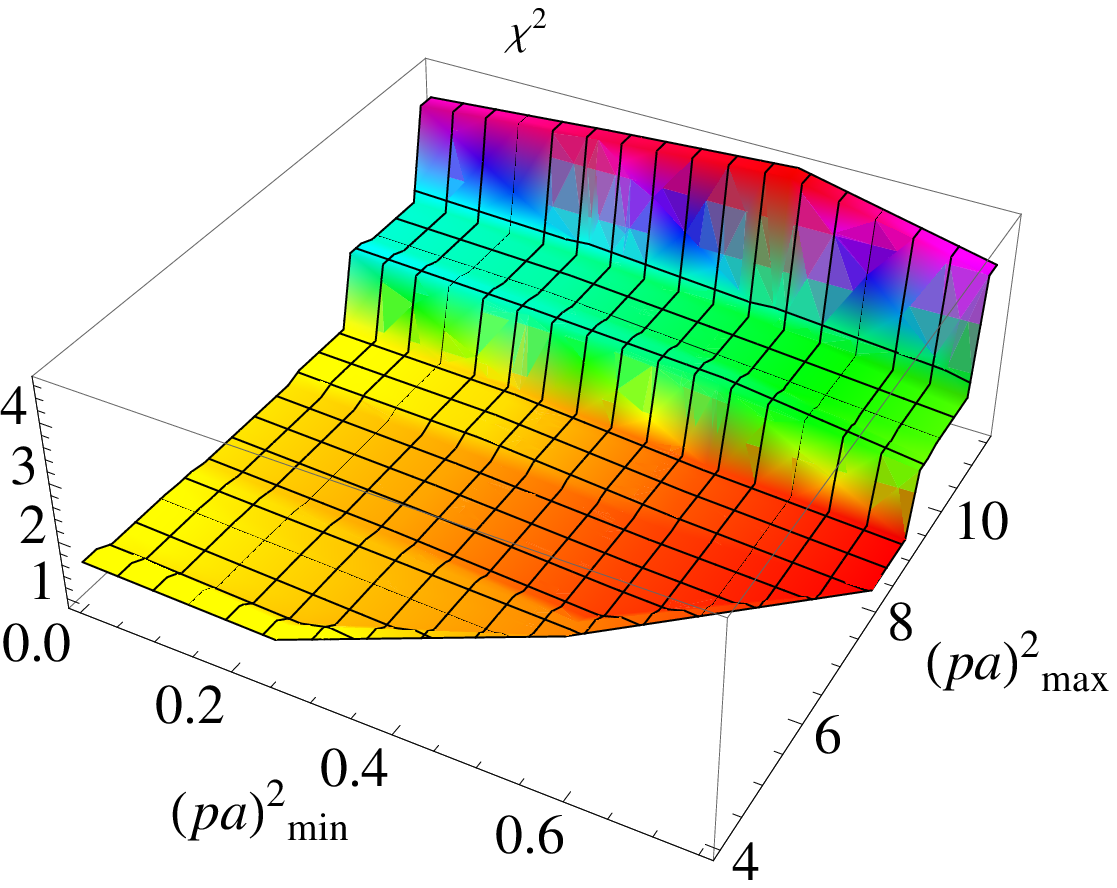}
&
       \includegraphics[scale=0.63,clip=true]{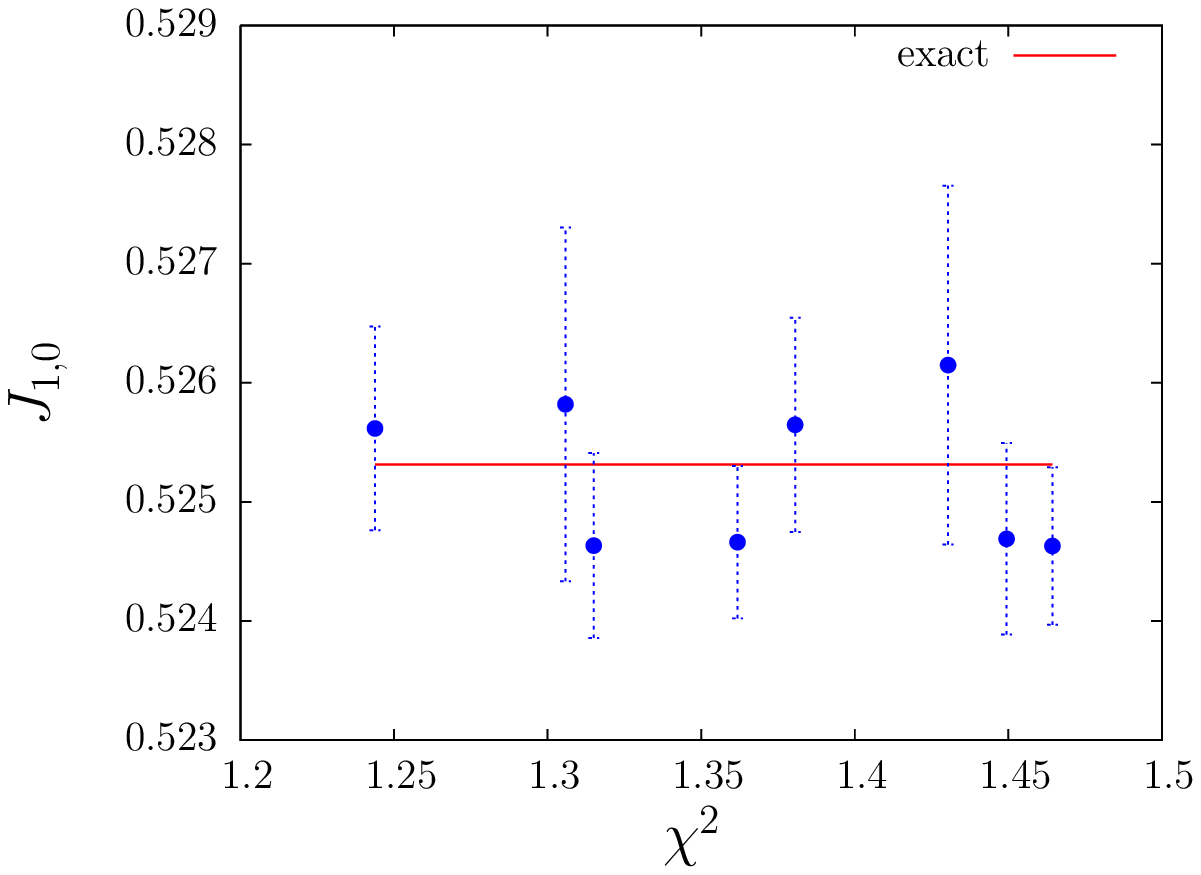}
    \end{tabular}
  \end{center}
  \caption{One-loop results. Left:  $\chi^2$
         as function of the fitting interval $\left[(pa)^2_{\rm min},(pa)^2_{\rm max}\right]$. 
         Right: $J_{1,0}$ for the smallest available 
         $\chi^2\gtrsim 1$ together with the exact value 0.525314 (see (\ref{J1loop})).}
  \label{fig:OneLoopFit}
\end{figure}
In Figure~\ref{fig:OneLoopFit} (left) we show the 
error $\chi^2$  as functions of $(pa)^2_{\rm min}$ and $(pa)^2_{\rm max}$
for the one-loop case. As discussed in the example of Section \ref{subsec:fitting}
the plot contains data for lattice sizes $N=8, 10, 12, 14, 16, 20$.
One recognizes several plateaus for the fit error $\chi^2$.
They prevent the determination of one optimal $\chi^2$ together with its attached fit parameters.
But rather they suggest to average over the smallest possible $\chi^2$.
In Figure~\ref{fig:OneLoopFit} (right) we show the values of the constant $J_{1,0}$ for the
smallest $\chi^2\gtrsim 1$ which agree within errors. 

Averaging over these data we obtain $J_{1,0}=0.52523(95)$. We notice that this value 
is stable with respect to the introduction of weights proportional to the inverse of $\chi^2$. 
The same holds for two and three loops.  

Another requirement for an efficient fit is a suitable relation of number of data points to
the number of fit parameters. In Table~\ref{tab:fitparnum1loop} this can be read off
for all considered orders.
\begin{table}[!htb]
\begin{center}
\begin{tabular}{|c|c|c|c|}
\hline
  $\chi^2$ & $N_{fp}$   & $N_{dp}$ & $J_{k,0}$\\ 
\hline
one-loop & & &\\
\hline
    1.244 & 13 & 43 & 0.52562(85)\\
    1.306 & 12 & 37 & 0.52582(148)\\
    1.315 & 11 & 31 & 0.52463(77)\\
    1.362 & 14 & 49 & 0.52466(64)\\
    1.381 & 15 & 55 & 0.52565(90)\\
    1.430 & 14 & 49 & 0.52615(150)\\
    1.449 & 13 & 43 & 0.52469(80)\\
    1.464 & 16 & 61 & 0.52463(66)\\
\hline
& & average : & 0.52523(95)\\
\hline
two-loop & & &\\
\hline
    2.769 & 11 & 31 & 1.4883(52)\\
    2.901 & 10 & 25 & 1.4910(87)\\
    2.941 & 14 & 49 & 1.4865(51)\\
    3.057 & 13 & 43 & 1.4864(74)\\
    3.064 & 16 & 61 & 1.4852(38)\\
    3.105 & 13 & 43 & 1.4860(40)\\
    3.123 & 12 & 37 & 1.4878(13)\\
    3.146 & 15 & 55 & 1.4855(66)\\
\hline
 & & average:& 1.4872(57)\\
\hline
three-loop  & & &\\
\hline
    3.869 & 12 & 25 & 5.04(19)\\
    4.117 & 11 & 21 & 4.66(35)\\
    4.455 & 11 & 21 & 5.02(23)\\
    4.485 & 13 & 29 & 5.14(18)\\
    4.753 & 12 & 25 & 4.83(31)\\
    4.913 & 16 & 41 & 5.23(17)\\
    5.084 & 10 & 17 & 4.62(48)\\
    5.104 & 15 & 37 & 5.03(28)\\
\hline
& & average:  & 4.94(27)\\
\hline
\end{tabular}
\end{center}
\caption{Number of fit parameters ($N_{fp}$) and data points ($N_{dp}$) for the  
$\chi^2$ taken to compute 
the constants $J_{k,0}$.}
\label{tab:fitparnum1loop}
\end{table}
\begin{figure}[!htb]
  \begin{center}
    \begin{tabular}{cc}
       \includegraphics[scale=0.63,clip=true]{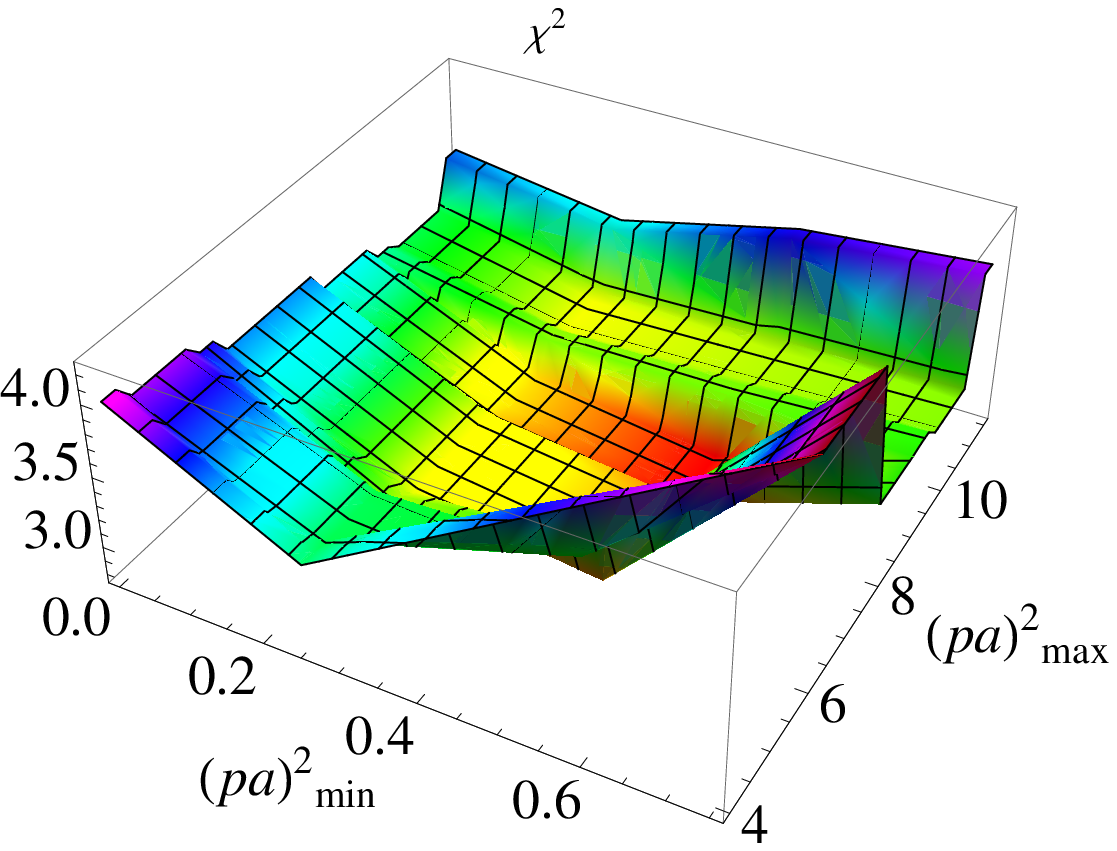}
&
       \includegraphics[scale=0.63,clip=true]{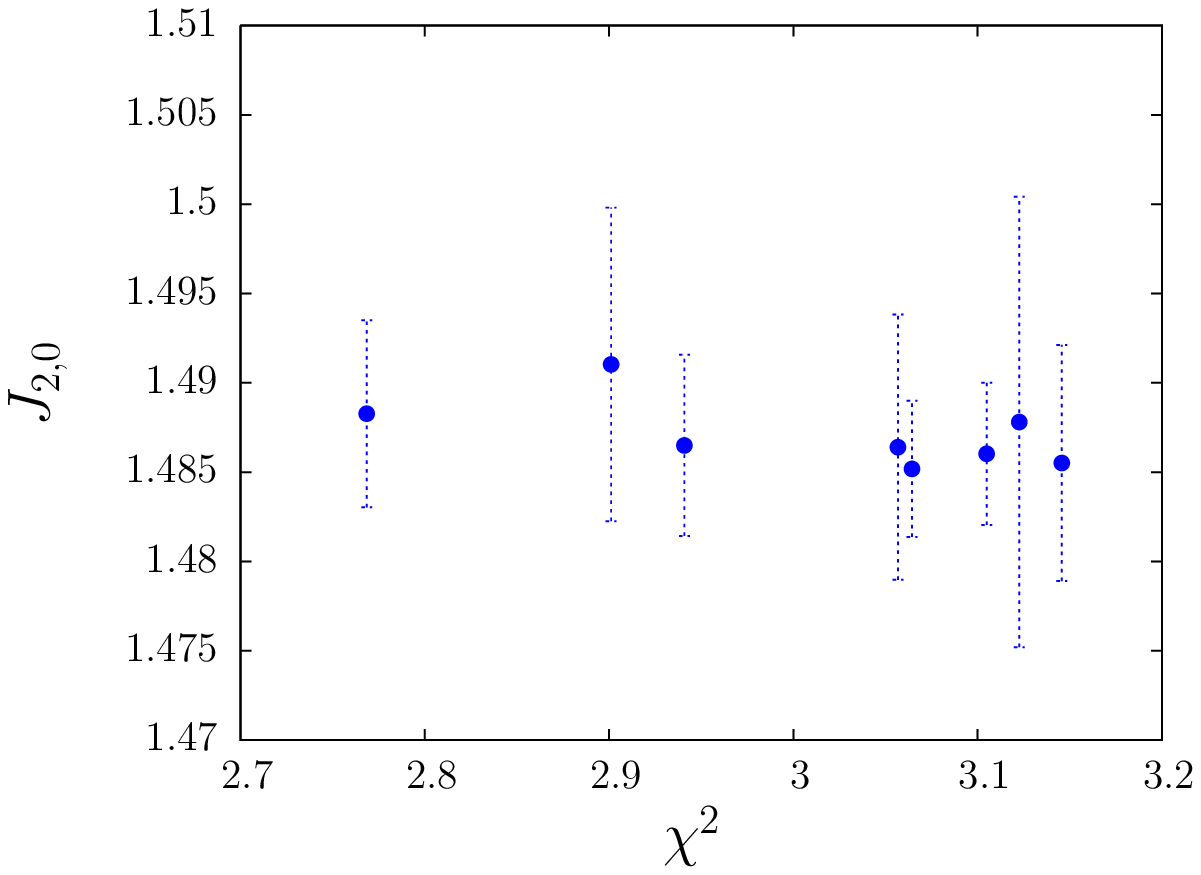}
    \end{tabular}
  \end{center}
  \caption{Same as in Figure~\ref{fig:OneLoopFit} for the two-loop results.}
  \label{fig:TwoLoopFit}
\end{figure}
The value $J_{1,0}$ almost perfectly coincides with the exactly known constant given in 
(\ref{J1loop}) and the estimated error is small.
This is a  clear indication that we have reached a sufficient good accuracy
of the accumulated data in the NSPT approach
and the proposed fitting procedure for handling both $O(pa)$ and $O(pL)$ effects works well. 
For two-loop order the corresponding results are shown in Figure~\ref{fig:TwoLoopFit}. 
Averaging over the eight best $\chi^2$ values we get $J_{2,0}=1.4872(57)$.

\begin{figure}[!htb]
  \begin{center}
    \begin{tabular}{cc}
       \includegraphics[scale=0.63,clip=true]{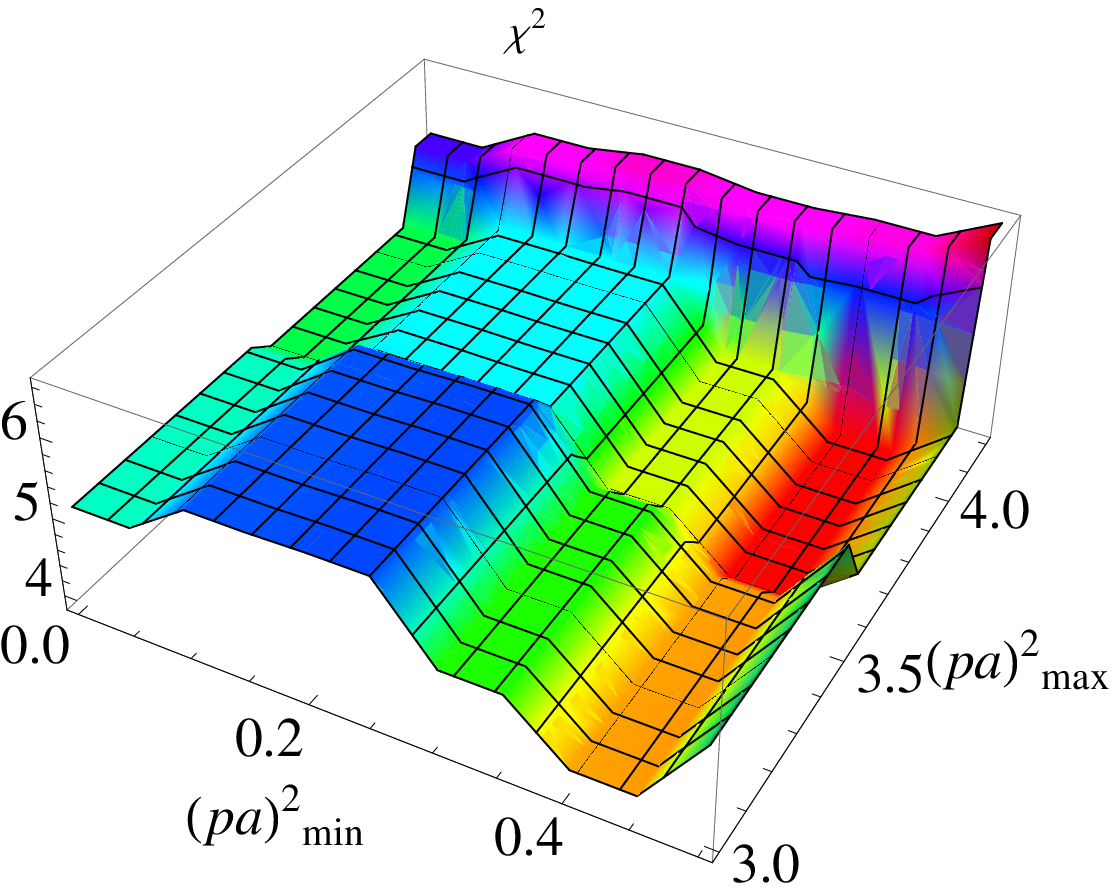}
&
       \includegraphics[scale=0.63,clip=true]{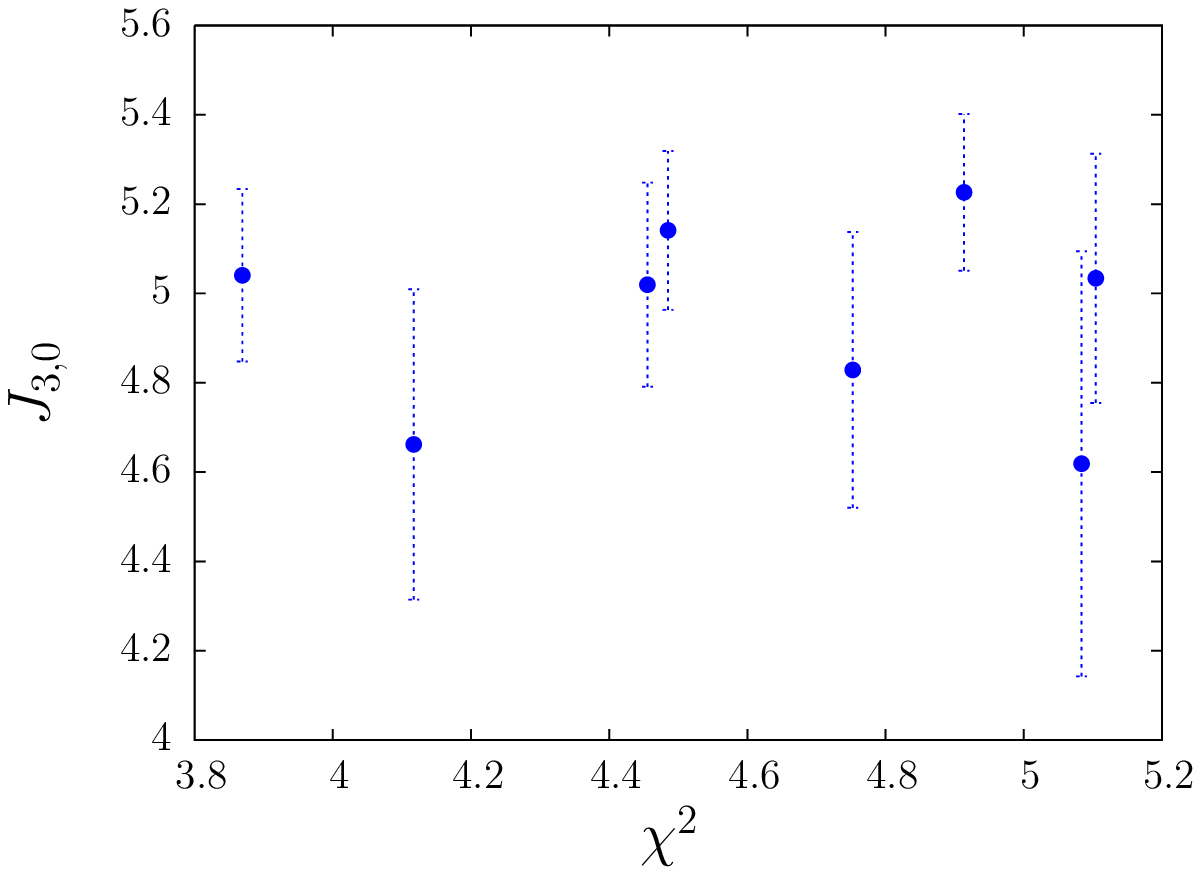}
    \end{tabular}
  \end{center}
  \caption{Same as in Figure~\ref{fig:OneLoopFit} for the three-loop results.}
  \label{fig:ThreeLoopFit}
\end{figure}

Three-loop data are available for lattice sizes $L=12, 14, 16, 20$ only. Therefore, the fits
have to be performed with a smaller set of input numbers. From Table~\ref{tab:fitparnum1loop}
one can deduce that the three-loop fit has the least accuracy.
The  results are shown in Figure~\ref{fig:ThreeLoopFit}. 
The average yields $J_{3,0}=4.94(27)$ which is supported by the Figure.

A major source of indetermination in our three loop analysis comes from the extrapolation 
$\varepsilon \to 0$. 
For three sizes ($N=14,16,20$) we only have four values of $\varepsilon$.
The number given above is obtained from a quadratic fit $\varepsilon \to 0$
used for one- and two-loop cases as well.
Computing $J_{3,0}$ from a linear $\varepsilon$-fit and using otherwise the same
strategy we obtain $5.40(11)$ which differs about $10\%$ from the quadratic
extrapolation. On the contrary, the one- and two-loop constants are
practically not affected by the two extrapolations.
For the linear extrapolation we get $J_{1,0}=0.52451(74)$ and $J_{2,0}=1.4867(43)$.

Collecting all results we can write (\ref{Zghostbeta3loop}) in a numerical form 
\begin{eqnarray}
&&J(a,p,\beta) = 1 + \frac{1}{\beta}\,\bigl( -0.0854897\, \log (pa)^2 + 0.52523(95) \bigr)
 + \nonumber \\
& &\hspace{-1cm} 
 + \frac{1}{\beta^2}\,\bigl( 0.0215195\, \left( \log (pa)^2 \right)^2 - 0.358423 \, 
\log (pa)^2 + 1.4872(57) \bigr) +
\label{J3loopNum} \\
& & \hspace{-1cm}
 + \frac{1}{\beta^3}\,\bigl( -0.0066027\, \left( \log (pa)^2 \right)^3 + 0.175434 \, 
\left( \log (pa)^2 \right)^2 - 1.6731 \, \log (pa)^2 + 4.94(27) \bigr)\,.
\nonumber
\end{eqnarray}
In (\ref{J3loopNum}) the coefficients  $J_{i,j}\, (j<i)$ are given with their significant 
digits only. 

Formulas (\ref{J2loop}) and (\ref{J3loop}) encode the relation to the $z^{\rm RI'}_{2,0}$ and 
$z^{\rm RI'}_{3,0}$ coefficients. Due to numerical accidents, these are affected by relevant 
percental errors.

\subsection{A first comparison to MC data}

Using the $A=-{\rm log} \,  U$ definition (\ref{eq:log_mapping}) as in NSPT in the
gluon propagator and the Faddeev-Popov operator (\ref{eq:FPop})
for the calculation of the ghost propagator, the Berlin Humboldt 
University group has obtained Monte Carlo results for the 
two propagators in Landau gauge and the resulting running
gauge coupling~\cite{Menz}. Maximizing the ``linear'' gauge 
fixing functional $\sum_{x,\mu} \Re \mathrm{tr} U_{x,\mu}$ as 
a preconditioner, the final gauge fixing was organized to minimize
\begin{equation}
\frac{1}{V} \sum_x {\rm Tr} \left[
    \left(\sum_\mu  \partial_\mu^L A_\mu \right)^\dagger (x)
    \left( \sum_\mu \partial_\mu^L A_\mu \right)(x) \right] 
\label{eq:fullcondition}
\end{equation}
with an iterative scheme following (\ref{eq:localstep}), actually 
in a parallelized multigrid realization of Fourier acceleration
(\ref{eq:fourieracc}).

Since it is assumed that non-perturbative contributions dominate 
mainly the intermediate and infrared momentum range,
it is of interest here to compare directly the perturbative ghost 
dressing function obtained in NSPT with its Monte Carlo counterpart 
for each common momentum 4-tuple.

We calculate the perturbative dressing function at a given lattice volume
summed up to loop order $n_{\rm {max}}$ for a
given lattice coupling $\beta$ as follows:
\begin{equation}
  \hat J= 1 + \sum_{n=1}^{n_{\rm {max}}} \frac{1}{\beta^n}
    \, \hat J^{(n)}\,.
\end{equation}

\begin{figure}[!htb]
   \begin{center}
   \includegraphics[scale=0.70,clip=true] {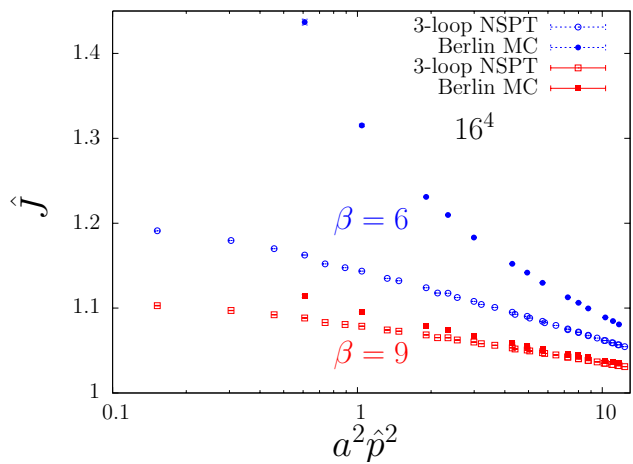}
   \end{center}
\caption{Three-loop NSPT of the ghost dressing function
in comparison to Monte Carlo at two $\beta$ values.}
\label{fig:5}
\end{figure}

In Figure~\ref{fig:5}
we compare the perturbative ghost dressing function at lattice size  $N=16$ 
with Monte Carlo data at two different $\beta$ values.
We observe that
at least three-loop accuracy is necessary to guarantee that the perturbative
ghost propagator at larger $\beta$ ($\beta \gtrsim 9.0$)
approximately describes
the full two-point function in the large momentum squared region
($p^2 > 1 {\rm~GeV}^2$).

\section{Summary}

We presented two- and three-loop results for the ghost propagator of Lattice SU(3) in
Landau gauge. A careful analysis of finite volume and finite lattice size effects has been performed, 
whose methodology has been discussed quite in detail. 
The present work will be supplemented by a similar analysis for the gluon propagator
(in the same gauge). This will open the way to a more careful comparison to Monte 
Carlo data for these quantities, which are 
supposed to encode informations on the confinement mechanism of 
non-Abelian gauge theories not only in the asymptotic infrared region,
but also in the intermediate momentum range in the form of power
corrections (condensates)~\cite{Boucaud:2008gn} and contributions from 
non-perturbative excitations 
(vortices etc.)~\cite{Langfeld:2001cz,Gattnar:2004bf} related to particular 
length scales.

\section*{Acknowledgements}
This work is supported by DFG under contract SCHI 422/8-1, DFG SFB/TR 55
and by I.N.F.N. under the research project MI11.
We acknowledge computer time made available to us by ECT$^*$ on the Ben system for 
approximately 13000 CPUh.

\end{document}